%% file: brexit_threat.tex
\documentclass[12pt]{article}
\usepackage[utf8]{inputenc}
\usepackage{textcomp}
\usepackage{amsmath}
\usepackage{graphicx}
\usepackage{fancyhdr}
\usepackage{lastpage}
\usepackage{natbib}
\usepackage{t1enc}

\usepackage[a4paper, top=25mm,bottom=25mm,left=25mm,right=25mm]{geometry}
\usepackage{afterpage}
\usepackage{amssymb}
\usepackage{subcaption}
\usepackage{enumerate}
\usepackage[usenames,dvipsnames,table]{xcolor}
\newcolumntype{L}{>{\raggedright\arraybackslash}X}
\newcolumntype{C}{>{\centering\arraybackslash}X}
\usepackage{booktabs}
\usepackage{tabularx}
\usepackage{nccmath}
\usepackage{multirow}
\usepackage{pgfplots}
\usepackage[hyphens]{url}
\pgfplotsset{compat=1.14}
\usepackage{tikz}
\usetikzlibrary{arrows,decorations.markings}
\pgfdeclarelayer{background}
\pgfsetlayers{background,main}
\usepackage{caption}
\usepackage{hyperref}
\hypersetup{
  colorlinks   = true,    
  urlcolor     = blue,    
  linkcolor    = blue,    
  citecolor    = blue      
}
\usepackage{authblk}
\usepackage{blindtext}
\usepackage{amsthm}
\theoremstyle{definition}
\newtheorem{definition}{Definition}

\theoremstyle{definition}
\newtheorem{example}{Example}
\usepackage{array}
\usepackage{dcolumn}
\usepackage{pdflscape}

\begin{document}

\title{Brexit: The Belated Threat}

\author[1]{D\'{o}ra Gréta Petr\'{o}czy}
\author[2]{Mark Francis Rogers}
\author[3]{Lászl\'{o} Á. K\'{o}czy}

\affil[1]{Department of Finance, 
Corvinus University of Budapest 

\texttt{doragreta.petroczy@uni-corvinus.hu}}
\affil[2]{Fazekas Mihály Secondary School of Budapest

\texttt {mark.francis.rogers@icloud.com} }
\affil[3]{Centre for Economic and Regional Studies,
Hungarian Academy of Sciences
and
Department of Finance,
Budapest University of Technology and Economics 

\texttt {koczy@krtk.mta.hu} }

\maketitle 

\begin{abstract}
Debates on an EU-leaving referendum arose in several member states after Brexit. We want to highlight how the exit of an additional country affects the power distribution in the Council of the European Union. We inspect the power indices of the member states both with and without the country which might leave the union. Our results show a pattern connected to a change in the threshold of the number of member states required for a decision. An exit that modifies this threshold benefits the countries with high population, while an exit that does not cause such a change benefits the small member states. According to our calculations, the threat of Brexit would have worked differently before the entry of Croatia.

\end{abstract}
 \providecommand{\keywords}[1]
{
  \small	
   \textbf{\textit{Keywords---}} #1
 }

\section{Introduction}\label{Sec1}
Since the European Union membership referendum of the United Kingdom (UK) in 2016, Brexit (the withdrawal of the United Kingdom from the European Union (EU)), its possible effects have become a subject of political debates in several countries, like the Czech Republic, France, or Greece \citep{the_guardian_brexit_frexit}. Although it might be worth inspecting several political and economic effects of an exit from the European Union, in this paper, we look at one aspect: how the power distribution changes in the Council of the European Union.
 \citet{koczy_how_2016} has shown that Brexit will mainly benefit large countries. In this paper, we firstly try to explore whether the same will remain true if another country will left.
Secondly, we want to answer the question: what would be the effect of Brexit if Croatia had not joined the EU?


The Council of the European Union, often referred to as the Council of Ministers, is an institute that represents the governments of the member states. It accepts EU law and synchronizes the policy of the EU. Along with the European Parliament, the Council of the European Union is the main decision-making body of the EU. Every member state is represented by an individual. The difference in size among the member states appears in a weighted qualified majority voting. Pursuant to the Treaty of Lisbon, a voting is successful if 
\begin{enumerate}
 \item at least 55\% of the member states (member quota), which
 \item represent at least 65\% of the habitants (population quota)
\end{enumerate}
support the decision. 
Such a recreation of the weights enables us to calculate how the power distribution changes if a country leaves the European Union. 

Several studies have addressed how voting power affects the overall likelihood of decision-making \citep{felsenthal_weighted_1997,felsenthal_treaty_2001}. \citet{wrantjen_votes_2017} has shown empirically that there is a stable positive relationship between the number of votes backing a member state request to change European legislation and its success probability. Therefore, it is an important question to measure how much power the countries have in the Council of the European Union.

Concerning our methodology, we use two well-known power indices: (1)  the Shapley-Shubik index  \citep{shapley_method_1954}; and (2) the Banzhaf index \citep{banzhaf_j._f._weighted_1965,coleman_control_1971,penrose_elementary_1946}. These measures show that if a decision is made, what probability a particular player has in being instrumental in making that decision. Translate this into voting about the spending of the budget, the indices reflect the probability of spending one (or a million) euro is according to the interests of this particular player. 
For several cases of departure, we show the change made by an exit until 2030, which can be called a 'farsighted' sense.  

We find a pattern connected to a change in the threshold of the required number of member states. An exit which changes in the population quota benefits the large, an exit that does not cause such a change benefits the small countries. Our results point in the direction that if the UK had left the European Union before the entry of Croatia, the effect would have been reversed, it would have favored the power of the small countries.

The paper is structured in the following way. Power indices to be used are defined and presented in Section \ref{Sec2}. The results of our calculations together with their interpretation are detailed in Section \ref{Sec3}.

\section{Methodology}\label{Sec2}

It is popular to study voting situations as simple cooperative games, where the players are the voters. The value of any coalition (a subset of the player set) is 1 if its players can decide a question, or 0 if not.
According to \citet{felsenthal_priori_2004}, there are two interpretations of voting power. One conception, the influence power (I-power) focuses on voting power conceived of as a voter's potential impact on the result of divisions of the decision-making institute: whether the policies proposed are adopted or rejected. The second conception, prize power (P-power) focuses on a voter's expected share of a fixed prize given to the winning coalition. It is more appropriate to use P-power if one wants to calculate the part of the EU budget a member can expect to control, however, it is not suitable to compare different voting situations.

The most common measures of power in voting games are the \emph{Shapley-Shubik}  and the \emph{Banzhaf indices}. They are used extensively for determining power in the Council of the European Union \citep{felsenthal_treaty_2001,herne_distribution_1993,koczy_beyond_2012, widgren_voting_1994}. Since we investigate a phenomenon that belongs to the P-Power, it is best to focus more on analyzing the power distribution of the Council of the European Union with the Shapley-Shubik index \citep{felsenthal_voting_1998}. 
 
The Shapley-Shubik index is an application of the Shapley value \citep{shapley_value_1953} for simple voting games. 
Its principle can be described as follows: voters arrive in a random order, and when a coalition becomes winning, the full credit is given to the \textit{pivotal} player arriving last. A player's power is specified by the proportion of orders in which it plays this role. 
The index shows that if a decision is made, what probability a particular player has in being instrumental in making that decision. 
  Let $N$ denote the set of players and let $S \subseteq N$ be an arbitrary subset of $N$. We use the corresponding lower-case letters to denote the cardinality of sets, so that $s= \lvert S \lvert $ and $n=\lvert N \lvert$. 

\begin{definition} (\emph{Shapley-Shubik index})
For any simple voting game $v$, player $i$'s \emph{Shapley-Shubik index} is as follows:
\[ \sum_{S  \subseteq N \setminus  \{i\}}  \frac{s!(n-s-1)!}{n!} \left(v\big(S\cup\left\{i\right\}\big)-v\big(S\big)\right). \] 
 \end{definition}
 
The Banzhaf index, which is the normalized Banzhaf-value \citep{banzhaf_j._f._weighted_1965,coleman_control_1971,penrose_elementary_1946}, uses a different approach.
A player is called \textit{critical} if it can turn a winning coalition into a losing one. The index shows what is the probability that a player influences a decision.

\begin{definition}
Player i's \emph{Banzhaf value} is:
\[ \sum_{S  \subseteq N \setminus  \{i\}}  \frac{1}{2^{n-1} }\left(v\big(S\cup\left\{i\right\}\big)-v\big(S\big)\right)= \frac{\eta_i (v)}{2^{n-1}}, \]
where  $\eta_i(v)$ is player $i$'s Banzhaf score, the number of coalitions where $i$  is critical.
\end{definition}

Usually its normalized value is reported as the measure of voting power.

\begin{definition} The \emph{Banzhaf index} is the normalized Banzhaf-score:
\[\beta_i =\frac{\eta_i (v)}{\sum_{j\in N} \eta_j (v)}
.\]

\end{definition}

 The index shows the voter's expected relative share of the total payoff.

When a country leaves, we assume that its payment to the EU budget ceases, therefore the others do not share the same prize as before.\footnote{~It is a simplification, as some non-EU member countries, like Norway, also contribute to the EU budget in a certain sense. The final financial conditions of the Brexit are unknown.} 
Taking this into account, we correct the power index by the following fraction:  
\begin{equation}\label{correction ratio}
\frac{\text{Original budget -- the payment of the leaving country}}{\text{Original budget}}.
\end{equation}

We compute for every country and each exit the adjusted power index as a percentage of the pre-exit power index. Since the adjusted power indices may not sum to 1, we cannot define them as probabilities, however, they still express a reasonable financial share. 
The following example presents an application of the new indices.
\begin{table}[!htbp]
\centering
\begin{tabularx}{1\linewidth}{lCCC} \toprule
Member State & Weight & S-S index  (\%) & Bz index (\%)\\ \midrule
France & 4                         & 23.33             &    23.80                                                     \\
Germany   & 4                         & 23.33               & 23.80                                                      \\
Italy   & 4                         & 23.33               & 23.80                                                      \\ \midrule
Belgium       & 2                         & 15.00                & 14.29                                                     \\
Netherlands     & 2                         & 15.00                 & 14.29                                                   \\ \midrule
Luxemburg     & 1                         & 0                   & 0                                                   \\  \bottomrule
\end{tabularx}
\caption{Decision-making in the Council of Ministers in 1958 \citep{koczy2009}, Shapley-Shubik (S-S) and Banzhaf (Bz) indices}
\label{tabl1}
\end{table}
\begin{example}\label{example1}
In the predecessor of the EU (the European Economic Community (ECC)), the six founding states already used a weighted voting system. The weight of the large countries (France, Germany, Italy) was 4, the weight of the medium-sized states (Belgium, the Netherlands) was 2, and the weight of the smallest state (Luxembourg) was 1. The decision threshold was 12.
\end{example}

According to Table \ref{tabl1}, Luxembourg's power was 0.
France, Germany, and Italy contributed 28\%  to the EEC budget, Belgium and the Netherlands paid 7.9\%, while Luxembourg paid only 0.2\%.
\begin{table}[!htbp]
\centering
\begin{tabularx}{1\linewidth}{lCCCC} \toprule
\multirow{2}{*}{Member state}      & S-S index \linebreak After   & Bz index \linebreak After   & Adjusted \linebreak S-S index  & Adjusted \linebreak Bz index  \\ \midrule
France & 23.33     & 23.80                  & 23.25   &23.32                                                                \\
Germany   & 23.33      & 23.80                  & 23.25                                                         &23.32        \\
Italy   & 23.33    & 23.80                    & 23.25    &23.32                                                                \\ \midrule
Belgium       & 15.00    & 14.29                    & 14.97  &14.00                                                                   \\
Netherlands     & 15.00   & 14.29                     & 14.97      &14.00                                                              \\ 
\bottomrule
\end{tabularx}
\caption{The effect of Luxembourg's departure from the Council of Ministers in 1958, Shapley-Shubik (S-S) and Banzhaf (Bz) indices in percentages}
\label{tabl_luxemburg}
\end{table}
If Luxembourg would have exited and the decision-threshold had not changed, the remaining countries' Shapley-Shubik and Banzhaf indices would have remained the same, but the adjusted indices would have decreased (see Table \ref{tabl_luxemburg}).

\begin{table}[!ht]
\centering
\begin{tabularx}{1\linewidth}{LCCCCCC} \toprule
Member \linebreak State      & S-S  index \linebreak  Before   & S-S  index\linebreak After    & Adjusted \linebreak S-S index & Bz index \linebreak Before   & Bz index \linebreak  After  & Adjusted \linebreak Bz  index \\ \midrule

Germany   & 23.33                       & 30.00              & $\downarrow$  21.60  & 23.80                       & 30.43               &  $\downarrow$ 21.91                                                    \\
Italy   & 23.33                        & 30.00             & $\downarrow$ 21.60     & 23.80                        & 30.43             & $\downarrow$ 21.91                                                 \\ \midrule
Belgium       & 15.00                         & 13.33               & $\downarrow$ 9.60   & 14.29                         & 13.04                & $\downarrow$ 9.39                                          \\
Netherlands     & 15.00                         & 13.33                & $\downarrow$ 9.60    & 14.29                         & 13.04                 & $\downarrow$ 9.39                                             \\ \midrule
Luxemburg     & 0                         & 13.33                   & $\uparrow$ 9.60    & 0                         & 13.04                    & $\uparrow$  9.39                                                  \\  \bottomrule
\end{tabularx}
\caption{The effect of France's departure from the Council of Ministers in 1958, Shapley-Shubik (S-S) and Banzhaf (Bz) indices in percentages}
\label{tabl_fro}
\end{table}

If a large country, for example, France, departs, and the threshold decreases to 9, then the change is more spectacular. The correction ratio, according to formula \eqref{correction ratio}, is 0.72. Table \ref{tabl_fro} shows the power measured by the adjusted Shapley-Shubik and Banzhaf indices. The only winner of this exit is Luxembourg.

\section{Results} \label{Sec3}

In this section, we present our findings.
Currently, pursuant to the Treaty of Lisbon, a qualified majority voting is successful in the Council of the European Union if 
\begin{enumerate}
 \item at least 55\% of the member states (\emph{member quota}), which
 \item represent at least 65\% of the habitants (\emph{population quota})
\end{enumerate}
support the decision. Furthermore, a blocking minority must include at least four Council members, failing which the qualified majority shall be deemed attained \citep{Council_EU}. This condition can be called the \emph{blocking minority rule}.\footnote{~For further details about the blocking minority rule, please see Appendix \hyperref[app_E]{E}.} 

We use population projections for 2015, 2020 and 2030 from Eurostat \citep{eurostat_europop2013_2014}  and budget data from the European Parliament \citep{european_parliament_eu_2015} (See Appendix \hyperref[app_A]{A}). The software IOP-Indices of Power \citep{brauninger_t._and_t._koni_indices_2005} is used to calculate the Shapley-Shubik and Banzhaf indices. The IOP cannot handle large numbers, thus population data are entered in 100,000s that may have a marginal effect on the indices. For the sake of simplicity, we disregarded the blocking minority rule in the calculations of adjusted power indices,  which also have some minor effect (see Appendix \hyperref[app_E]{E}). 

\citet{koczy_how_2016} has shown that if the United Kingdom leaves the European Union, which have currently 28 member states, the smallest member states' power indices decrease. We have found the same result after repeating the calculation for every other member state (see Appendix \hyperref[app_B]{B}).
However, a further question remains: what happens if another member state leaves the EU? Here, we discuss the effects of the Czech Republic (Czexit) or Germany leaving the EU, after Brexit. 
Secondly, building on our previous finding, we have inspected what would be the effect of Brexit on the power distribution of the EU had the United Kingdom left it before Croatia entered? Is Brexit in this sense a belated threat? Our results show that it is. 

In the following, we will call a country large or small regarding its population. We observe a general pattern, which connects the change in the member state quota to a change in the power distribution: when this threshold is modified by the departure, the power indices of the large countries increase. When such a change is not evoked by the departure, the power indices of the small countries increase.
\subsection{The impact of additional departures to Brexit}

In the computations which investigate the results of an \textit{additional departure to Brexit}, we base our calculations on the 27-member Union without the UK, because Brexit appears to be a fact if any further exit happens. As mentioned in the previous section, it is also considered that an exit of a country causes a change in the budget. The example of the Czech Republic is presented first because the EU-skeptical sentiment has become stronger recently in this country.\footnote{~In January 2018, Miloš Zeman, the openly Eurosceptic President of the Czech Republic  was re-elected.} The budget correction ratio is 0.989 according to formula \eqref{correction ratio}. Figure \ref{Fig1} shows the budget-adjusted change in power indices due to Czexit as the function of the population.
\input{Figure1}
\input{Figure3}

We find that in the case of Czexit, the power indices of the small countries increase, and the power indices of the large countries such as France, Germany, Italy, Poland and Spain slightly decrease. The main winners from Czexit are Cyprus, Estonia,  Luxembourg and Malta.
\input{Figure4}
The same can be said if one investigates Czexit in a farsighted sense, meaning to repeat the analysis with population predictions for 2020 and 2030. The only country which power index change differs is Romania: from a slight decrease (see Figure \ref{Fig1}), its power modestly increases (see Figure \ref{Fig3}).

We get similar results for other departures from a 27-member EU: the power indices of the small countries increase significantly. The detailed results for all member states can be seen in Appendix \hyperref[app_C]{C}. 
What has created more variation in these cases is the contribution of the particular country to the EU budget. To illustrate this point, let us look at the exit of Germany.

In the case of Germany's exit (Figure \ref{germany_a}), the Shapley-Shubik indices of the smallest countries and Poland increase, while the other countries all lose power. This is because countries with large populations are also the ones that contribute the most, so the budget loss exceeds the power gains caused by the departure of Germany. The correction ratio \eqref{correction ratio}  is 0.711.

The results concerning Poland are especially interesting. If one of the four large countries (Germany, France, Italy, or Spain) leaves, Poland is much better off than Romania or Spain which are the closest countries in size of the population. In all four cases, its Shapley-Shubik index increases despite the other remaining large countries' power decrease.

The simulations have been repeated with the other popular power measure, the Banzhaf index. 
We get the same results, the power of small countries increases. The largest difference is in the case of Germany. As one can see in Figure \ref{germany_b}, with the use of the Banzhaf index, all countries, including Poland lose power. 
As there is no significant difference, and the Banzhaf index rather represents the I-Power approach \citep{felsenthal_priori_2004}, the Shapley-Shubik index is applied in the following.
\input{Figure6}

Calculations of another country leaving the 26-member EU, for instance, if the Czech Republic leaves after Germany show a similar pattern to Brexit (Figure \ref{Fig6}). This can be elucidated by the fact that as the number of member states decreases from 26 to 25, the Council of the European Union's threshold for the number of supporting member states (determined by the member quota) decreases from 15 to 14. In this case, small countries would lose while the power of the large countries would increase. 

\subsection{The effect of Brexit before the accession of Croatia}
Since our findings on an additional departure show an impact that is the inverse of Brexit's \citep{koczy_how_2016}, Brexit might have a different impact before the accession of Croatia compared to the exit from the 28-member EU.  

This has significance because if Brexit had decreased the power of large countries such as France and Germany, the impact of the potential Brexit would have been calculated differently by these states that usually dominate the policy of the EU: Brexit would have been a greater risk for them. 
In other words, if Brexit would have had the reverse impact before Croatia's membership, it could be seen as a belated threat.

We find that Brexit before the accession of Croatia would have favored smaller and would have harmed larger countries (Figure \ref{Fig7}). The results are similar not only for Brexit but for the case of an exit of any other member state from the EU without Croatia (see Appendix \hyperref[app_D]{D}).
\input{Figure7}


\input{Figure8}

\subsection{A generalization of the results}
Note that an additional departure to Brexit has an inverted impact compared to Brexit's impact from the 28-member EU, but it is similar to the potential effect of Brexit if it had happened before Croatia's membership. Results for a departure from the hypothetical 26-member European Union have a strong resemblance to the consequences of Brexit. 
 The inverted impact of an additional departure to Brexit is due to the fact that 15 countries are necessary to make a voting successful in the case of both 26 and 27 members. However, after an additional exit the population threshold decreases. 
 
The voting rule states two main requirements: the support of a given number of countries and a certain percentage of the population. A country will turn a losing coalition into a winning one if (a) the coalition just misses a member state to pass the threshold, and/or (b) if the coalition has the required participation, but the supporting countries are too small to reach the population quota. 

With Czexit after Brexit, the population threshold decreases while the member state threshold remains the same, so coalitions with smaller countries become winning, which shifts power from the large to the small member states. This pattern is quite prevalent, we find similar results using population projections for 2020 and 2030 (Figure \ref{Fig3}).

It seems to be a general pattern that an exit triggering a decrease in the quota benefits large, while an exit not triggering such a change benefits small member states.

In the case of 27 member states, voting is successful if at least 15 countries, having together at least a population of 288 million vote in favor. We have examined the number of countries whose power increases if a particular country leaves, which can be considered as a yes vote for the exit of the departing country. Figure \ref{Fig8} presents the number of countries and their total population with an increasing power.
Most of the countries would get a positive vote for leaving from 20 or 21 countries, but without the required population. However, in the case of Poland, both thresholds are met, because the power of small and large countries increase, and merely some medium countries lose power.

\section{Conclusions}

Inspired by Brexit, the goal of our investigation has been to examine what would happen in the Council of the European Union after a country's exit from the EU. For this purpose, the potential changes in the influence of each country have been measured with the use of power indices.

We find that, not just Brexit, but any other exit from the 28-member EU would favor countries with high population.  However, an additional exit would increase the power of small countries. Furthermore, we observe a general pattern which is linked to the change in the member-state threshold. An exit, which changes the number of member states required for a decision, benefits the large, while an exit that does not cause such a change benefits the small countries. Thus a hypothetical Brexit before the accession of Croatia would harm large countries' power in the Council.



\bibliographystyle{SageH}
\small\bibliography{allbib_eng}
\newpage

\section*{Appendix A -- Population and contribution data}\label{app_A}
\addcontentsline{toc}{section}{Appendix A}
\input{Appendix_A}

\newpage
\section*{Appendix B -- The impact of any member state leaving the 28-member EU}\label{app_B}
\addcontentsline{toc}{section}{Appendix B}
The following table presents the impact of any member state leaving the 28-member EU. The country labels in the columns refer to the country that is leaving the EU, the rows show the remaining member states. The values represent the change (new adjusted S-S power index)/(old adjusted S-S power index) in basis points (1/100th of 1\%).  Bold indicates incresing, while italic signs decreasing power.\\
\input{Appendix_B.tex}
\pagebreak
\clearpage
\section*{Appendix C -- The impact of additional departures to Brexit}\label{app_C}
\addcontentsline{toc}{section}{Appendix C}
The following table presents the impact of any member state leaving the 27-member EU, after the United Kingdom departs. The country labels in the columns refer to the country that is leaving the EU, the rows show the remaining member states. The values represent the change (new adjusted S-S power index)/(old adjusted S-S power index) in basis points (1/100th of 1\%). Bold indicates increasing, while italic signs decreasing power.\\

\input{Appendix_C}
\clearpage
\section*{Appendix D -- The impact of any member state leaving before the accession of Croatia}\label{app_D}
\addcontentsline{toc}{section}{Appendix D}
The following table presents the impact of any member state leaving the 27-member EU before the accession of Croatia. The country labels in the columns refer to the country that is leaving the EU, the rows show the remaining member states. The values represent the change (new adjusted S-S power index)/(old adjusted S-S power index) in basis points (1/100th of 1\%). Bold indicates increasing, while italic signs decreasing power.\\

\input{Appendix_D_alternative}

\clearpage
\section*{Appendix E -- The blocking minority rule}\label{app_E}
\addcontentsline{toc}{section}{Appendix E}
According to the Article 16(4) of the Treaty on European Union \emph{''as from 1 November 2014, a qualified majority shall be defined as at least 55\% of the members of the Council, comprising at least fifteen of them and representing Member States comprising at least 65\% of the population of the Union.
A blocking minority must include at least four Council members, failing which the qualified majority shall be deemed attained.''}\footnote{\url{https://eur-lex.europa.eu/resource.html?uri=cellar:2bf140bf-a3f8-4ab2-b506-fd71826e6da6.0023.02/DOC_1&format=PDF} }

For the sake of simplicity, we left out the blocking minority rule in the calculations of the adjusted power indices with the use of the software IOP.
In the following, the effect of this modification will be calculated.

In the current 28-member state case, there are only 10 variants of coalitions that are winning only due to the blocking minority rule. Table \ref{bm_28} shows all coalitions that are not blocking minorities even though they reach the population quota. 
\begin{table}[!ht]

\centering
\begin{tabularx}{0.9\linewidth}
{cCCC} 

1 & Germany & France &  United Kingdom \\
2 & Germany & France & Italy \\
3 & Germany & France & Spain \\
4 & Germany & France & Poland \\
5 & Germany & United Kingdom & Italy \\
6 & Germany & United Kingdom & Spain \\
7 & Germany & United Kingdom & Poland \\
8 & Germany & Italy & Spain \\
9 & Germany & Italy & Poland \\
10 & France & United Kingdom & Italy \\
\end{tabularx}
\caption{Coalitions which reach the population quota but cannot reject a decision in the 28-member EU}
\label{bm_28}
\end{table}

In the case of small countries, in other words, for countries not appearing in Table  \ref{bm_28} (their number is 23), by ignoring the blocking minority rule, we do not take them as a pivotal player in 10 possible variations, but they are.
Thus their Shapley-Shubik index should be increased by $(24!\cdot 3!\cdot 10)/28!={1}/{8190}=0.000122$.

In the case of France, Germany, Italy, Poland, Spain and the United Kingdom we need to reduce the index.
If France, Italy and the United Kingdom oppose a decision, they cannot block it, until another country joins them, so Germany is not considered as a pivotal player despite it plays this role. At the same time, we have counted Germany in nine variants as a pivotal player (for example, in the blocking coalition of France, Germany and the United Kingdom), but it does not play such a role. Therefore, the correction for Germany is:
\[\frac{24!\cdot3!-25!\cdot2!\cdot9}{28!}=-\frac{444}{491400}=-0.000904.\]

After Brexit, in the 27-member EU, there are 27! possible coalitions, and 19 variants involved in the correction needed due to the blocking minority rule.
\begin{table}[!ht]

\centering
\begin{tabularx}{0.8\linewidth}
{cCCC} 
1& Germany& France& Italy\\
2& Germany& France& Spain\\
3& Germany& France& Poland\\
4& Germany& France& Romania\\
5& Germany& France& Netherlands\\
6& Germany& France& Belgium\\
7& Germany& France& Greece\\
8& Germany& France& Czech Republic\\
9& Germany& France& Portugal\\
10& Germany& France&Hungary\\
11& Germany& France&Sweden\\
12& Germany& France&Austria\\
13& Germany& Italy& Spain\\
14& Germany& Italy& Poland\\
15& Germany& Italy& Romania\\
16& Germany& Italy& Netherlands \\
17& Germany& Spain& Poland\\
18& France& Italy& Spain\\
19& France& Italy& Poland \\

\end{tabularx}
\caption{Coalitions that reach the population quota but cannot reject a decision in the 27-member EU after Brexit}
\label{bm_27}
\end{table}

By ignoring the blocking minority rule, in the case of countries not appearing in Table \ref{bm_27} (their number is 12), we do not take them as a pivotal player in 19 possible variants despite the fact that they are.
Their Shapley-Shubik index should be increased by $(23!\cdot 3!\cdot 19)/27!={19}/{70200}=0.000271$.

We show the overall effect of these corrections for Malta. The Shapley-Shubik index of Malta, calculated by the IOP software without the blocking minority rule, is 0.008487, which needs to be increased by $1/8190$. After Brexit, the Shapley-Shubik index  of Malta according to our calculation is 0.008036. As mentioned, it should be increased by 19/70200. With the payment correction, the adjusted Shapley-Shubik index will be 0.007574. Therefore the accurate change in power is $0.007574/(0.008487 + 0.000122) = 0.879751$. The original result was 0.863331, the difference is only 0.016421. Since Malta has the smallest Shapley-Shubik value, the adjustment for the other countries is lower. Consequently, ignoring the blocking minority rule does not have any significant effect on our results.
\end{document}

%% file: Figure1.tex
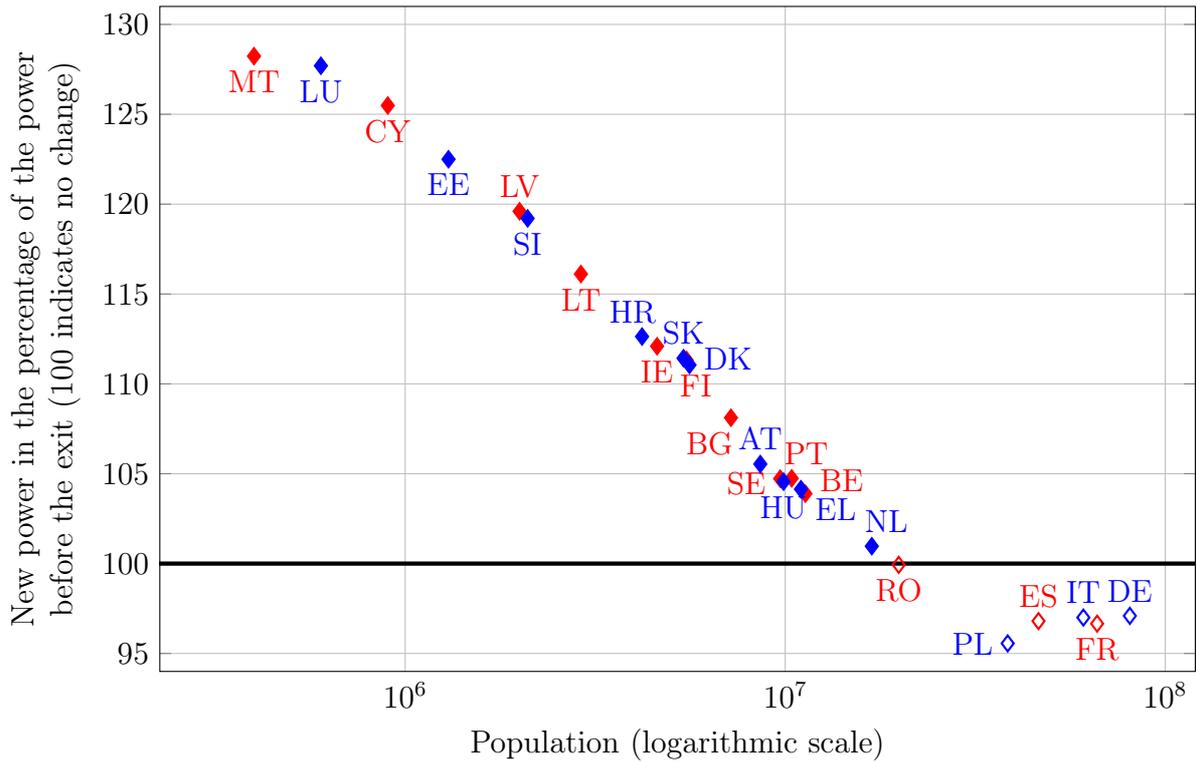
\begin{figure}[!ht]
\centering

\begin{tikzpicture}
\begin{axis}[width=0.95\textwidth, 
height=0.65\textwidth,
xlabel = Population (logarithmic scale),
ylabel = New power in the percentage of the power \\ before the exit ($100$ indicates no change),
ylabel style={align=center},
xmax=120000000,
ymin=94,
ymax=131,
xmode = log,
xtick style={draw=none},
xmajorgrids,
ymajorgrids]

\addplot[
only marks,
thick,
mark=diamond,
mark size = 3,
red,
visualization depends on=\thisrow{alignment} \as \alignment,
nodes near coords,
point meta=explicit symbolic,
every node near coord/.style={anchor=\alignment,inner sep=5pt},
] 
table [meta index=2] {
x	y	label	alignment
19900000	99.93009361	RO	90
46400000	96.80117378	ES	270
66200000	96.65830624	FR	90
};

\addplot[
only marks,
thick,
mark=diamond,
mark size = 3,
blue,
visualization depends on=\thisrow{alignment} \as \alignment,
nodes near coords,
point meta=explicit symbolic,
every node near coord/.style={anchor=\alignment,inner sep=5pt},
] 
table [meta index=2] {
x	y	label	alignment
38500000	95.55904018	PL	0
60900000	96.99785938	IT	270
80700000	97.09127297	DE	270
};

\addplot[
only marks,
thick,
mark=diamond*,
mark size = 3,
red,
visualization depends on=\thisrow{alignment} \as \alignment,
nodes near coords,
point meta=explicit symbolic,
every node near coord/.style={anchor=\alignment,inner sep=5pt},
] 
table [meta index=2] {
x	y	label	alignment
400000	128.2332103	MT	90
900000	125.4887736	CY	90
2000000	119.6015019	LV	270
2900000	116.1111577	LT	90
4600000	112.0941157	IE	90
5500000	111.3318564	FI	110
7200000	108.1143507	BG	50
9700000	104.7260943	SE	8
10400000	104.7432613	PT	240
11300000	103.8916601	BE	200
};

\addplot[
only marks,
thick,
mark=diamond*,
mark size = 3,
blue,
visualization depends on=\thisrow{alignment} \as \alignment,
nodes near coords,
point meta=explicit symbolic,
every node near coord/.style={anchor=\alignment,inner sep=5pt},
] 
table [meta index=2] {
x	y	label	alignment
600000	127.6984076	LU	90
1300000	122.4970026	EE	90
2100000	119.2056736	SI	90
4200000	112.6325635	HR	290
5400000	111.4261347	SK	270
5600000	111.0526673	DK	190
8600000	105.5383693	AT	270
9900000	104.5582386	HU	90
11000000	104.1405268	EL	150
16900000	100.9677467	NL	240
};

\draw [ultra thick] (axis cs:\pgfkeysvalueof{/pgfplots/xmin},100)  -- (axis cs:120000000,100);
\end{axis}
\end{tikzpicture}

\caption{Effect of Czexit with populations for 2015, adjusted Shapley-Shubik index}
\label{Fig1}
\end{figure}

%% file: Figure3.tex
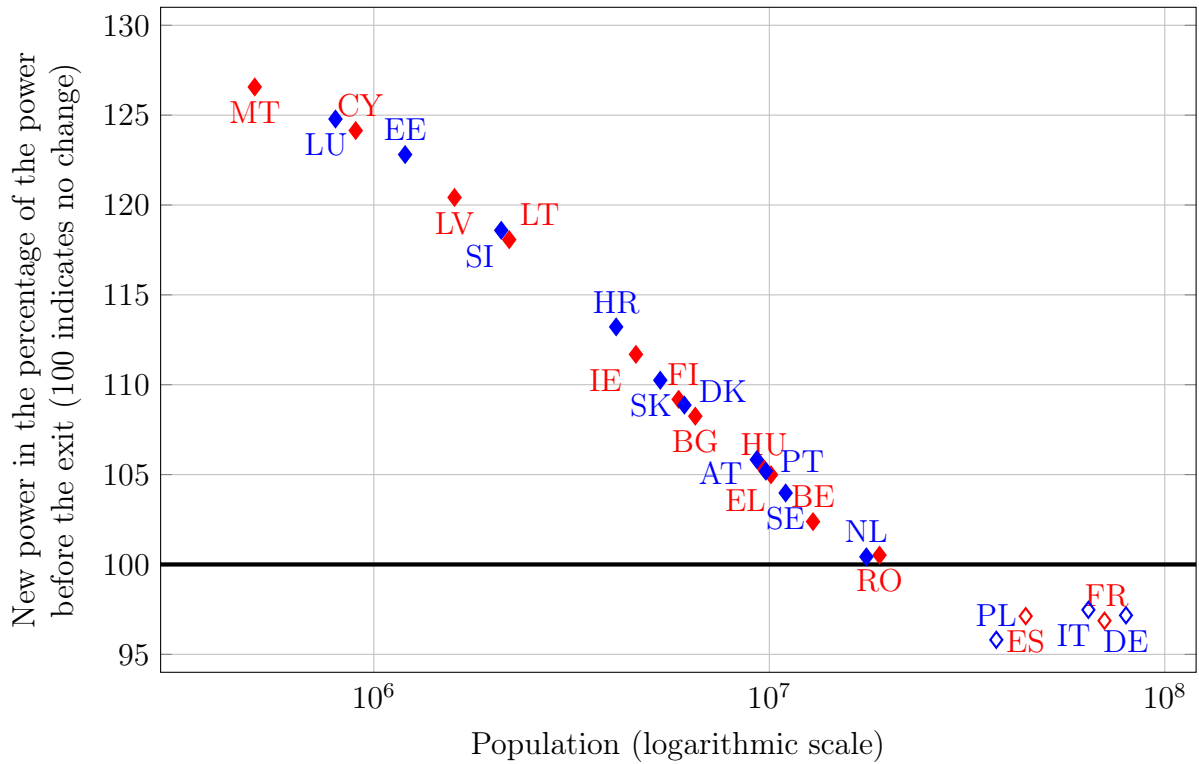
\begin{figure}[!ht]
\centering

\begin{tikzpicture}
\begin{axis}[width=0.95\textwidth, 
height=0.65\textwidth,
xlabel = Population (logarithmic scale),
ylabel = New power in the percentage of the power \\ before the exit ($100$ indicates no change),
ylabel style={align=center},
xmax=120000000,
ymin=94,
ymax=131,
xmode = log,
xtick style={draw=none},
xmajorgrids,
ymajorgrids]

\addplot[
only marks,
thick,
mark=diamond,
mark size = 3,
red,
visualization depends on=\thisrow{alignment} \as \alignment,
nodes near coords,
point meta=explicit symbolic,
every node near coord/.style={anchor=\alignment,inner sep=5pt},
] 
table [meta index=2] {
x	y	label	alignment
44500000 97.1252974395074 ES 90
70400000 96.8684318605574 FR 265

};

\addplot[
only marks,
thick,
mark=diamond,
mark size = 3,
blue,
visualization depends on=\thisrow{alignment} \as \alignment,
nodes near coords,
point meta=explicit symbolic,
every node near coord/.style={anchor=\alignment,inner sep=5pt},
] 
table [meta index=2] {
x	y	label	alignment
37500000 95.8070351553913 PL 270
64100000 97.4757398042339  IT 60
79800000 97.1695915795361 DE 90

};

\addplot[
only marks,
thick,
mark=diamond*,
mark size = 3,
red,
visualization depends on=\thisrow{alignment} \as \alignment,
nodes near coords,
point meta=explicit symbolic,
every node near coord/.style={anchor=\alignment,inner sep=5pt},
] 
table [meta index=2] {
x	y	label	alignment
500000	126.566181895459		MT 90
900000	124.140954190927		CY 260
1600000	120.419125495464		LV 90
2200000	118.069758581251		LT 220
4600000	111.688951617843		IE 40
5900000	109.185892046177		FI 260
6500000	108.250722685407		BG 90
9700000	105.316332525724		HU 270
10100000 104.990685093339	EL 45
12900000 102.383451263762	BE 270
19000000 100.51755223833		RO 90

};

\addplot[
only marks,
thick,
mark=diamond*,
mark size = 3,
blue,
visualization depends on=\thisrow{alignment} \as \alignment,
nodes near coords,
point meta=explicit symbolic,
every node near coord/.style={anchor=\alignment,inner sep=5pt},
] 
table [meta index=2] {
x	y	label	alignment
800000	124.782725594134		LU 70
1200000	122.805903754346		EE 270
2100000	118.590721528801		SI 50
4100000	113.217203606139		HR 270
5300000	110.24906103826	 SK 70
6100000	108.866744639944		DK 200
9300000	105.830938489083		AT 20
9800000	105.191992870334		PT 195
11000000	103.978364332652		SE 90
17600000	100.427401314486		NL 270

};

\draw [ultra thick] (axis cs:\pgfkeysvalueof{/pgfplots/xmin},100)  -- (axis cs:120000000,100);
\end{axis}
\end{tikzpicture}

\caption{Effect of Czexit with population projections for 2030, adjusted Shapley-Shubik index}
\label{Fig3}
\end{figure}

%% file: Figure4.tex
\begin{figure*}[!htbp]
\centering
\begin{subfigure}[h]{1\textwidth}
\caption{Adjusted Shapley-Shubik index}
\begin{tikzpicture}
\begin{axis}[width=0.95\textwidth, 
height=0.65\textwidth,
xlabel = Population (logarithmic scale),
ylabel = New power in the percentage of the power \\ before the exit  ($100$ indicates no change),
ylabel style={align=center},
xmax=120000000,
ymin=78,
ymax=104,
xmode = log,
xtick style={draw=none},
xmajorgrids,
ymajorgrids]

\addplot[
only marks,
thick,
mark=diamond,
mark size = 3,
red,
visualization depends on=\thisrow{alignment} \as \alignment,
nodes near coords,
point meta=explicit symbolic,
every node near coord/.style={anchor=\alignment,inner sep=5pt},
] 
table [meta index=2] {
x	y	label	alignment

900000	99.6100452713732	CY	90
2000000	96.8504252828442	LV	270
2900000	95.1487464272432	LT	270
4600000	93.2601169280716	IE	250
5500000	92.5321312881201	FI	90
7200000	91.1597906734613	BG	270
9700000	89.627460029241	 SE	90
10400000	89.5995382267589	PT	270
11000000	89.4319346478122	EL	135
16900000	88.5294409735091	NL	210
60900000	96.1668138644069	IT	90

};

\addplot[
only marks,
thick,
mark=diamond,
mark size = 3,
blue,
visualization depends on=\thisrow{alignment} \as \alignment,
nodes near coords,
point meta=explicit symbolic,
every node near coord/.style={anchor=\alignment,inner sep=5pt},
] 
table [meta index=2] {
x	y	label	alignment
1300000	98.2299027101594	EE	90
2100000	96.6676361981848	SI	90
4200000	93.6460052686549	HR	20
5400000	92.5332826928737	SK	210
5600000	92.3621689870996	DK	20
8600000	89.6186166936651	AT	20
9900000	89.6806589528901	HU	340
10500000	89.5673739358898	CZ	210
11300000	89.4033667938979	BE	180
19900000	88.1943867892189	RO	180
46400000	95.0495028431151	ES	90
66200000	97.5043110515662	FR	180

};

\addplot[
only marks,
thick,
mark=diamond*,
mark size = 3,
red,
visualization depends on=\thisrow{alignment} \as \alignment,
nodes near coords,
point meta=explicit symbolic,
every node near coord/.style={anchor=\alignment,inner sep=5pt},
] 
table [meta index=2] {
x	y	label	alignment

400000	101.024882217885	MT	270
38500000	102.013688756605	PL	270

};

\addplot[
only marks,
thick,
mark=diamond*,
mark size = 3,
blue,
visualization depends on=\thisrow{alignment} \as \alignment,
nodes near coords,
point meta=explicit symbolic,
every node near coord/.style={anchor=\alignment,inner sep=5pt},
] 
table [meta index=2] {
x	y	label	alignment

600000	100.513426160061	LU	270

};

\draw [ultra thick] (axis cs:\pgfkeysvalueof{/pgfplots/xmin},100)  -- (axis cs:120000000,100);
\end{axis}
\end{tikzpicture}

\label{germany_a}
\end{subfigure}

\begin{subfigure}[h]{1\textwidth}
\caption{Adjusted Banzhaf index}
\begin{tikzpicture}
\begin{axis}[width=0.95\textwidth, 
height=0.65\textwidth,
xlabel = Population (logarithmic scale),
ylabel = New power in the percentage of the power \\ before the exit ($100$ indicates no change),
ylabel style={align=center},
xmax=120000000,
ymin=78,
ymax=104,
xmode = log,
xtick style={draw=none},
xmajorgrids,
ymajorgrids]

\addplot[
only marks,
thick,
mark=diamond,
mark size = 3,
red,
visualization depends on=\thisrow{alignment} \as \alignment,
nodes near coords,
point meta=explicit symbolic,
every node near coord/.style={anchor=\alignment,inner sep=5pt},
] 
table [meta index=2] {
x	y	label	alignment

400000	94.1880175333804	MT	90
900000	93.791767515555	CY	270
2000000	92.9804514264766	LV	90
2900000	92.4507767635523	LT	270
4600000	91.48803415971	IE	290
5500000	91.0839051806949	FI	90
7200000	90.4421406752359	BG	290
9700000	89.7027225667645	SE	15
10400000	89.5415569644623	PT	60
11000000	89.4199356681038	EL	215
16900000	88.9861860069717	NL	140
38500000	86.5931703372437	PL	270
60900000	85.5994800433378	IT	320

};

\addplot[
only marks,
thick,
mark=diamond,
mark size = 3,
blue,
visualization depends on=\thisrow{alignment} \as \alignment,
nodes near coords,
point meta=explicit symbolic,
every node near coord/.style={anchor=\alignment,inner sep=5pt},
] 
table [meta index=2] {
x	y	label	alignment
600000	94.019338346343	LU	270
1300000	93.4713236271994	EE	90
2100000	92.9085875037197	SI	270
4200000	91.6802204158912	HR	20
5400000	91.1248785445325	SK	270
5600000	91.0433524339634	DK	20
8600000	89.9740605611201	AT	275
9900000	89.6600301028322	HU	245
10500000	89.5242506259291	CZ	175
11300000	89.3625684967632	BE	110
19900000	89.5511138514597	RO	270
46400000	79.7334393561242	ES	90
66200000	86.347203839819	FR	270

};



\draw [ultra thick] (axis cs:\pgfkeysvalueof{/pgfplots/xmin},100)  -- (axis cs:120000000,100);
\end{axis}
\end{tikzpicture}
\label{germany_b}
\end{subfigure}

\caption{Effect of the German exit}
\label{Fig4}
\end{figure*}
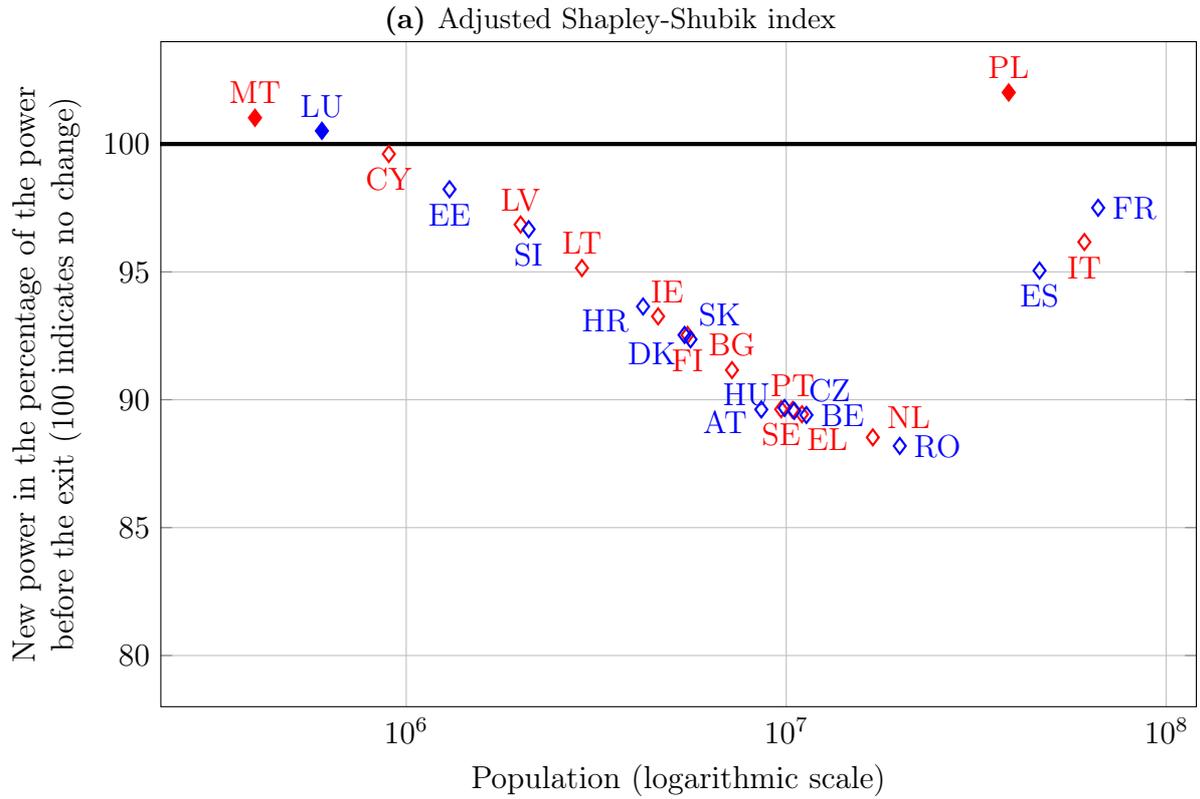
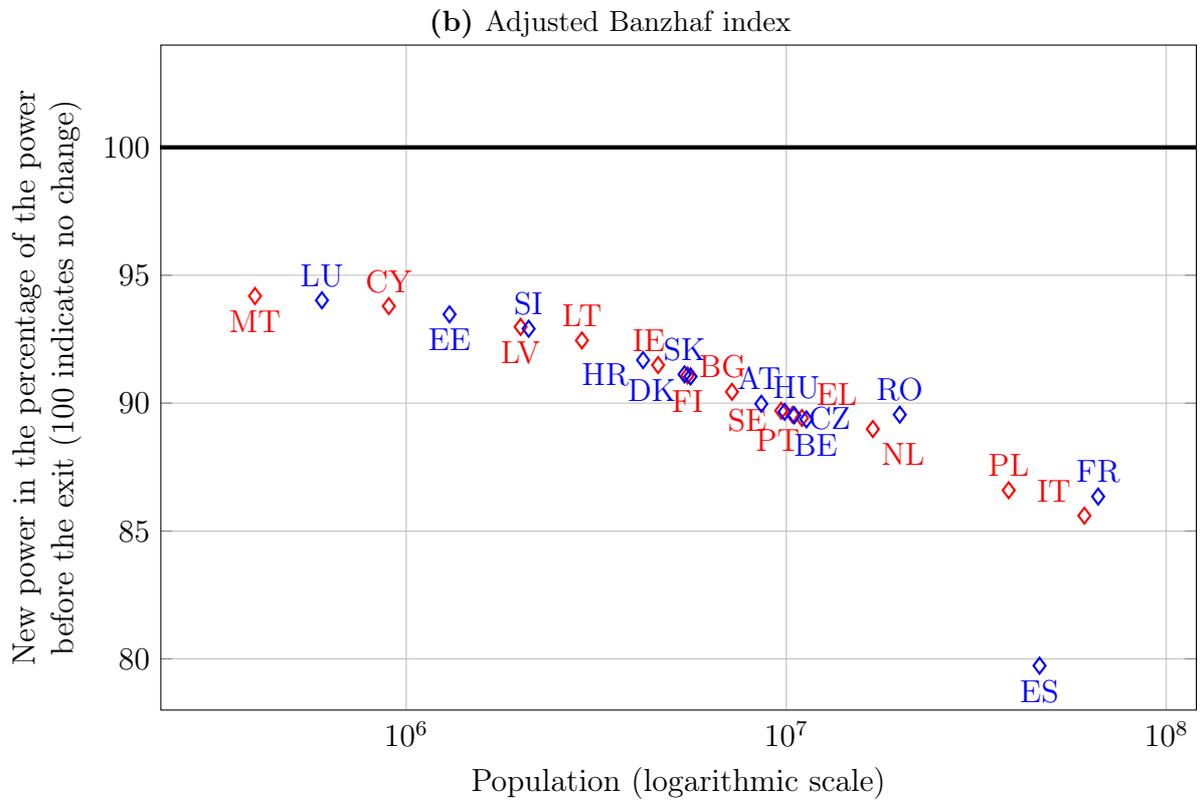

%% file: Figure6.tex
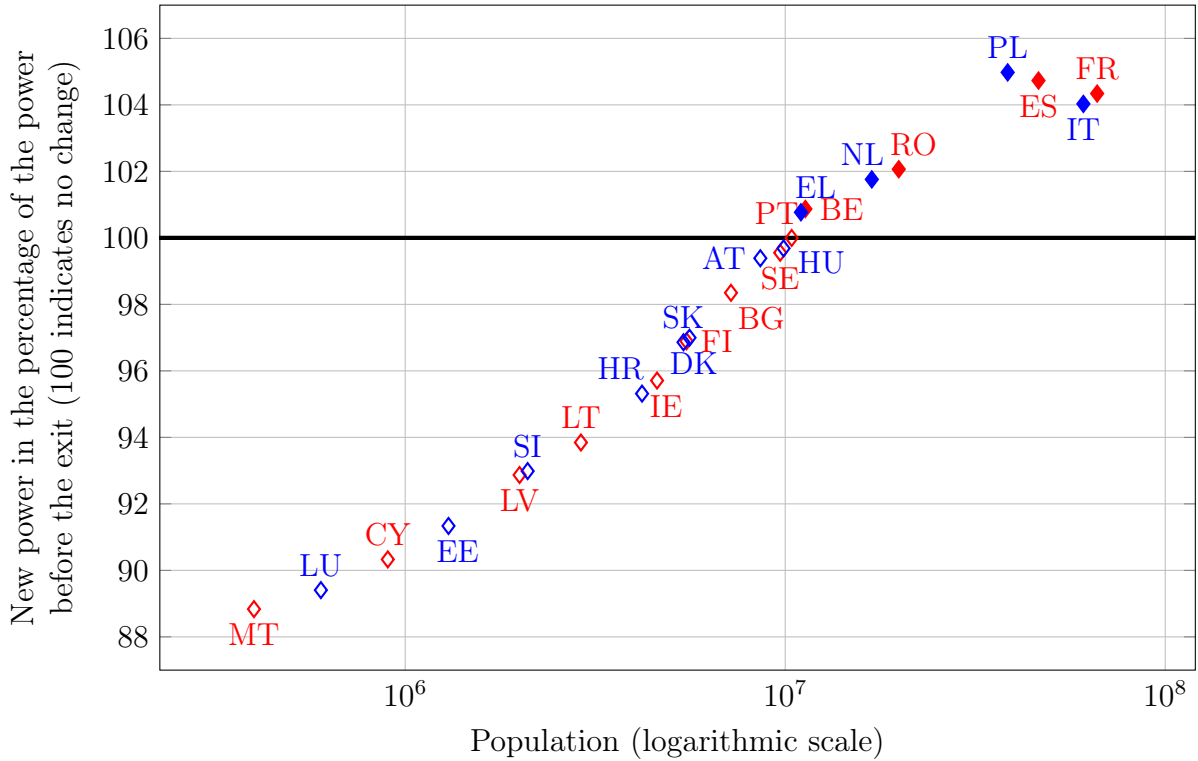
\begin{figure}[!ht]
\centering

\begin{tikzpicture}
\begin{axis}[width=0.95\textwidth, 
height=0.65\textwidth,
xlabel = Population (logarithmic scale),
ylabel = New power in the percentage of the power \\ before the exit ($100$ indicates no change),
ylabel style={align=center},
xmax=120000000,
ymin=87,
ymax=107,
xmode = log,
xtick style={draw=none},
xmajorgrids,
ymajorgrids]

\addplot[
only marks,
thick,
mark=diamond,
mark size = 3,
red,
visualization depends on=\thisrow{alignment} \as \alignment,
nodes near coords,
point meta=explicit symbolic,
every node near coord/.style={anchor=\alignment,inner sep=5pt},
] 
table [meta index=2] {
x	y	label	alignment

400000	88.8350220318513	MT	90
900000	90.3301559573276	CY	270
2000000	92.8685976149433	LV	90
2900000	93.8458607370396	LT	270
4600000	95.7088714482611	IE	110
5500000	96.8924280830273	FI	180
7200000	98.3474506116836	BG	140
9700000	99.5528947943834	SE	90
10400000	99.9969978946405	PT	300

};

\addplot[
only marks,
thick,
mark=diamond,
mark size = 3,
blue,
visualization depends on=\thisrow{alignment} \as \alignment,
nodes near coords,
point meta=explicit symbolic,
every node near coord/.style={anchor=\alignment,inner sep=5pt},
] 
table [meta index=2] {
x	y	label	alignment
600000	89.4062663284817	LU	270
1300000	91.3359964832341	EE	110
2100000	92.9859124416712	SI	270
4200000	95.3158167770217	HR	310
5400000	96.8602022660048	SK	270
5600000	96.9976627606139	DK	100
8600000	99.3870522412688	AT	0
9900000	99.6845132063506	HU	160

};

\addplot[
only marks,
thick,
mark=diamond*,
mark size = 3,
red,
visualization depends on=\thisrow{alignment} \as \alignment,
nodes near coords,
point meta=explicit symbolic,
every node near coord/.style={anchor=\alignment,inner sep=5pt},
] 
table [meta index=2] {
x	y	label	alignment

11300000	100.869797282059	BE	180
19900000	102.064141574558	RO	240
46400000	104.730813160001	ES	90
66200000	104.340183602144	FR	270

};

\addplot[
only marks,
thick,
mark=diamond*,
mark size = 3,
blue,
visualization depends on=\thisrow{alignment} \as \alignment,
nodes near coords,
point meta=explicit symbolic,
every node near coord/.style={anchor=\alignment,inner sep=5pt},
] 
table [meta index=2] {
x	y	label	alignment

11000000	100.769845492643	EL	240
16900000	101.758236864936	NL	290
38500000	104.976589104139	PL	270
60900000	104.029089145669	IT	90

};

\draw [ultra thick] (axis cs:\pgfkeysvalueof{/pgfplots/xmin},100)  -- (axis cs:120000000,100);
\end{axis}
\end{tikzpicture}

\caption{Change in power due to Czexit in the 26-member EU (after Brexit and Germany’s exit), adjusted Shapley-Shubik index}
\label{Fig6}
\end{figure}

%% file: Figure7.tex
\begin{figure}[!htbp]
\centering

\begin{tikzpicture}
\begin{axis}[width=0.95\textwidth, 
height=0.65\textwidth,
xlabel = Population (logaritmic scale),
ylabel = New power in the percentage of the power \\ before the exit ($100$ indicates no change),
ylabel style={align=center},
xmax=120000000,
ymin=89,
ymax=131,
xmode = log,
xtick style={draw=none},
xmajorgrids,
ymajorgrids]

\addplot[
only marks,
thick,
mark=diamond,
mark size = 3,
red,
visualization depends on=\thisrow{alignment} \as \alignment,
nodes near coords,
point meta=explicit symbolic,
every node near coord/.style={anchor=\alignment,inner sep=5pt},
] 
table [meta index=2] {
x	y	label	alignment
19900000	95.9400943266113	RO	90

};

\addplot[
only marks,
thick,
mark=diamond,
mark size = 3,
blue,
visualization depends on=\thisrow{alignment} \as \alignment,
nodes near coords,
point meta=explicit symbolic,
every node near coord/.style={anchor=\alignment,inner sep=5pt},
] 
table [meta index=2] {
x	y	label	alignment
16900000	98.6019064615173	NL	90
38500000	94.5541432691809	PL	270

};

\addplot[
only marks,
thick,
mark=diamond*,
mark size = 3,
red,
visualization depends on=\thisrow{alignment} \as \alignment,
nodes near coords,
point meta=explicit symbolic,
every node near coord/.style={anchor=\alignment,inner sep=5pt},
] 
table [meta index=2] {
x	y	label	alignment
400000	126.868742871285	MT	90
900000	123.736414698863	CY	270
2000000	118.860498794812	LV	270
2900000	115.306118695219	LT	270
5400000	109.261957915474	SK	90
5600000	109.06374886309	DK	270
8600000	105.238406628913	AT	25
9900000	103.69551144325	HU	255
10500000	103.142145492872	CZ	210
11300000	102.25811082841	BE	155
46400000	101.520083859199	ES	270
66200000	100.717605490308	FR	270

};

\addplot[
only marks,
thick,
mark=diamond*,
mark size = 3,
blue,
visualization depends on=\thisrow{alignment} \as \alignment,
nodes near coords,
point meta=explicit symbolic,
every node near coord/.style={anchor=\alignment,inner sep=5pt},
] 
table [meta index=2] {
x	y	label	alignment
600000	125.566070372449	LU	270
1300000	121.825790781068	EE	270
2100000	118.443688059013	SI	90
4600000	110.382143737621	IE	10
5500000	109.122501188519	FI	180
7200000	107.649423983865	BG	90
9700000	103.746799217223	SE	35
10400000	103.281027148715	PT	170
11000000	102.544657646853	EL	70
60900000	100.507787231771	IT	90
80700000	102.34058484968	DE	270

};

\draw [ultra thick] (axis cs:\pgfkeysvalueof{/pgfplots/xmin},100)  -- (axis cs:120000000,100);
\end{axis}
\end{tikzpicture}

\caption{Effect of Brexit before Croatia joined the EU, adjusted Shapley-Shubik index}
\label{Fig7}
\end{figure}
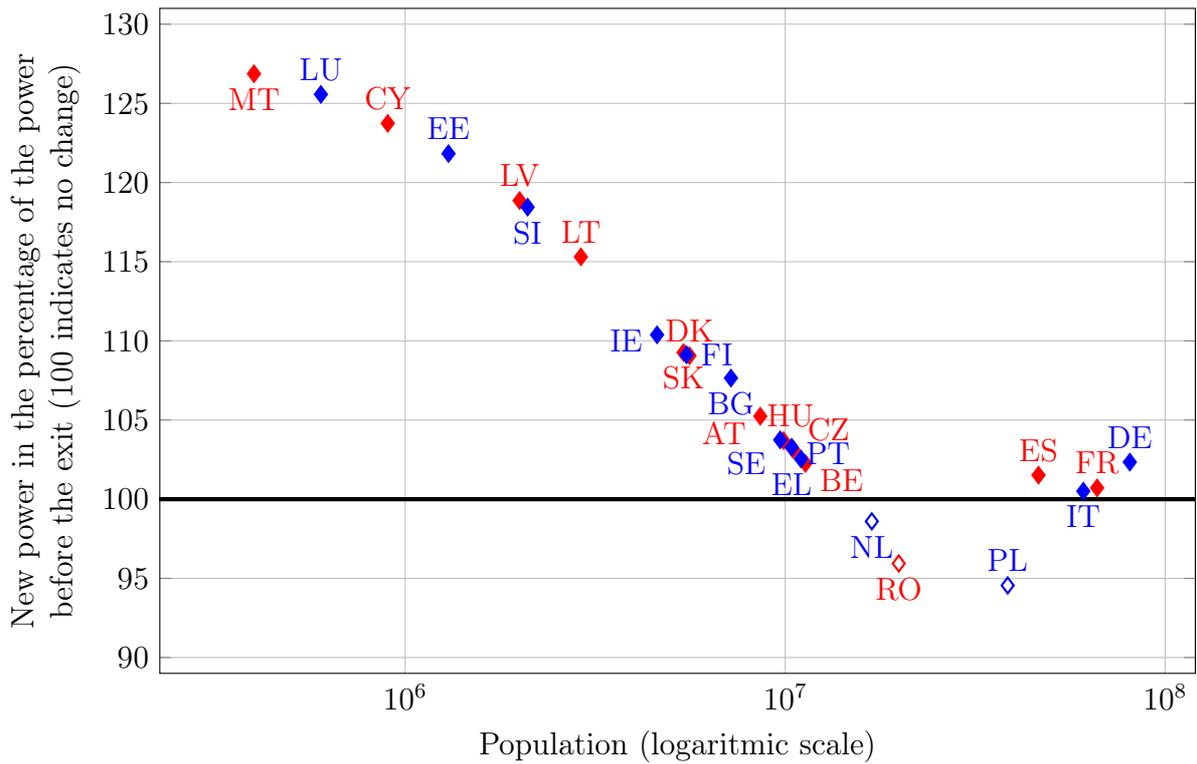

%% file: Figure8.tex
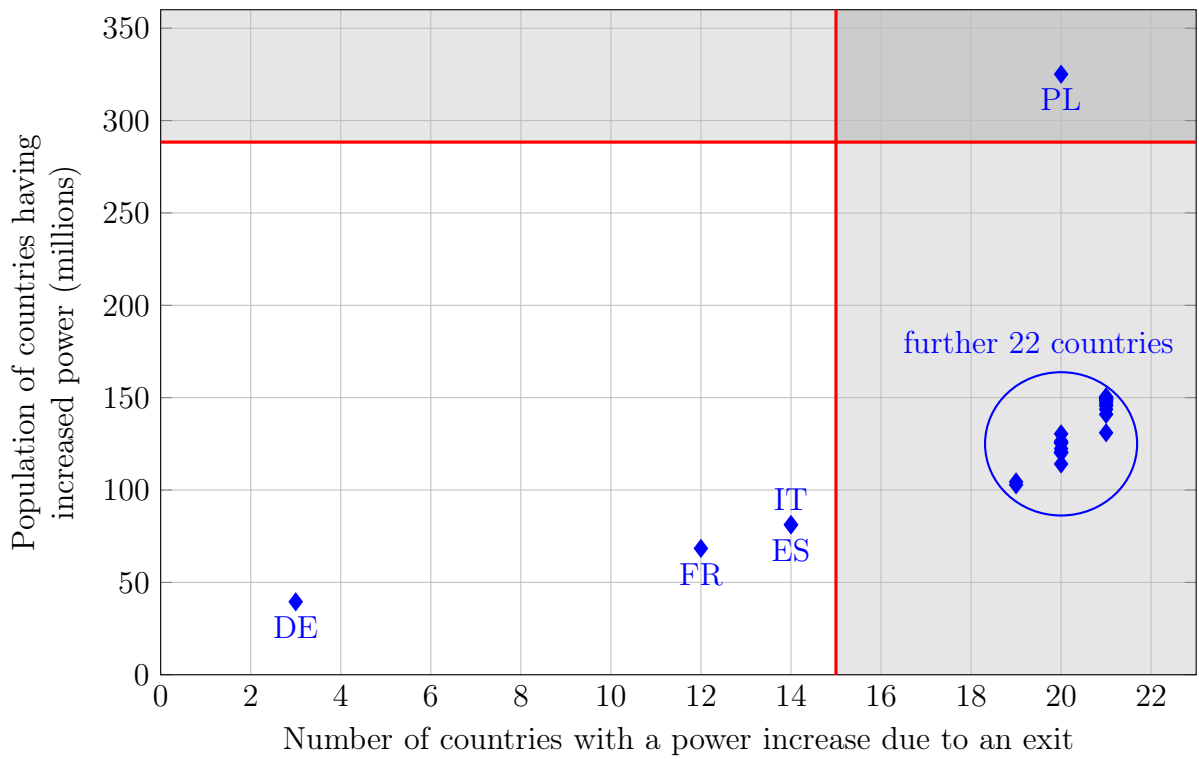
\begin{figure}[!htbp]
\centering

\begin{tikzpicture}
\begin{axis}[width=0.95\textwidth, 
height=0.65\textwidth,
/pgf/number format/.cd,
use comma,
xlabel =  Number of countries with a power increase due to an exit,
ylabel =  Population of countries having \\ increased power (millions),
ylabel style={align=center},
xmin=0,
xmax=23,
ymin=0,
ymax=360.000000,
xtick style={draw=none},
xmajorgrids,
ymajorgrids,
]

\addplot[
only marks,
thick,
mark=diamond*,
mark size = 3,
blue,
visualization depends on=\thisrow{alignment} \as \alignment,
nodes near coords,
point meta=explicit symbolic,
every node near coord/.style={anchor=\alignment,inner sep=5pt},
] 
table [meta index=2] {
x	y	label	alignment
3	39.500000	DE	90
12	68.400000	FR	90
14	81.200000	IT	270
14	81.200000	ES	90
19	102.800000	{}	90
19	104.400000	{}	90
20	122.400000	{}	90
20	120.500000	{}	90
20	125.400000	{}	90
20	125.500000	{}	90
20	120.000000	{}	90
20	126.400000	{}	90
20	130.400000	{}	90
20	114.100000	{}	90
20	325.100000	PL	90
20	120.600000	{}	90
21	143.700000	{}	90
21	146.700000	{}	90
21	150.000000	{}	90
21	149.600000	{}	90
21	141.000000	{}	90
21	148.900000	{}	90
21	148.000000	{}	90
21	150.500000	{}	90
21	131.000000	{}	90
21	145.500000	{}	90
21	148.800000	{}	90
};
\begin{pgfonlayer}{background}
\addplot+[fill,gray!40,no marks] coordinates {
(15,288.340000)
(15,360.000000)
(23,360.000000)
(23,288.340000)
};

\addplot+[fill,gray!20,no marks] coordinates {
(15,288.340000)
(15,0)
(23,0)
(23,288.340000)
};

\addplot+[fill,gray!20,no marks] coordinates {
(0,288.340000)
(0,360.000000)
(15,360.000000)
(15,288.340000)
};
\end{pgfonlayer}

\node [thick,blue] at (axis cs: 19.5,180.000000) {further $22$ countries};

\draw [thick,blue] (20,125.000000) ellipse (1cm and 0.95cm);

\draw [very thick,red] (0,288.340000) -- (23,288.340000);
\draw [very thick,red] (15,0) -- (15,360.000000);
\end{axis}
\end{tikzpicture}

\caption{Effect of a departure from the EU after Brexit, adjusted Shapley-Shubik index}
\label{Fig8}
\end{figure}

%% file: Appendix_A.tex
\begin{table}[!htbp]
\centering
\rowcolors{1}{gray!20}{}
\begin{tabularx}{\linewidth}{Lccccr} \toprule \hiderowcolors
\multicolumn{1}{L}{\begin{tabular}[l]{@{}l@{}}Member  \\ state \end{tabular}} & Abbrev. & \multicolumn{1}{c}{\begin{tabular}[c]{@{}c@{}}Population \\ 2015  \end{tabular}} & \multicolumn{1}{c}{\begin{tabular}[c]{@{}c@{}}Population \\ 2020 \end{tabular}}& \multicolumn{1}{c}{\begin{tabular}[c]{@{}c@{}}Population \\ 2030\  \end{tabular}} & \multicolumn{1}{c}{\begin{tabular}[c]{@{}c@{}}Contrib. \\ ratio (\%)
\end{tabular}} \\ \midrule \showrowcolors
Austria  &  AT &  86 &88&  93 & 1.22\\
Belgium  &  BE &  113 &118&  129 & 2.85\\
Bulgaria  &  BG &  72 &70&  65 & 0.31\\
Croatia  &  HR &  42 &42&  41 & 0.3\\
Cyprus  &  CY &  9 &9&  9 & 0.11\\
Czech Republic  &  CZ &  105 &106&  108 & 1.02\\
Denmark  &  DK &  56 &58&  61 & 1.72\\
Estonia  &  EE &  13 &13&  12 & 0.14\\
Finland  &  FI &  55 &56&  59 & 1.38\\
France  &  FR &  662 &677&  704 & 15.22\\
Germany  &  DE &  807 &806&  798 & 20.08\\
Greece  &  EL &  110 &107&  101 & 1.42\\
Hungary  &  HU &  99 &98&  97 & 0.69\\
Ireland  &  IE &  46 &46&  46 & 1.11\\
Italy  &  IT &  609 &620&  641 & 11.18\\
Latvia  &  LV &  20 &19&  16 & 0.19\\
Lithuania  &  LT &  29 &27&  22 & 0.25\\
Luxembourg  &  LU &  6 &6&  8 & 0.18\\
Malta  &  MT &  4 &4&  5 & 0.05\\
Netherlands  &  NL &  169 &171&  176 & 4.97\\
Poland  &  PL &  385 &384&  375 & 2.74\\
Portugal  &  PT &  104 &101&  98 & 1.27\\
Romania  &  RO &  199 &197&  190 & 1.05\\
Slovakia  &  SK &  54 &54&  53 & 0.49 \\
Slovenia  &  SI &  21 &21&  21 & 0.25 \\
Spain  &  ES &  464 &458&  445 & 7.76 \\
Sweden  &  SE &  97 &101&  110 & 2.98 \\
United Kingdom   &  UK &  646 &667&  705 & 8.82 \\
\hline   
\end{tabularx}
\caption{Member states of the EU -- population (in 100,000s) and financial contribution}
\label{appendixA}             
\end{table}

%% file: Appendix_B.tex

\begin{table}[!htbp]
\centering
\begin{footnotesize}
\centerline{
\rowcolors{1}{gray!20}{}
\begin{tabularx}{1.15\textwidth}{l CCCCC CCCCC CCCC}  \toprule \hiderowcolors
	 & AT & BE & BG & CY & CZ & DE & DK & EE & EL & ES & FI & FR & HR & HU \\ \hline
\showrowcolors
AT	&	\cellcolor{gray!80} &	\itshape $\downarrow$287 &	\itshape $\downarrow$124 &	\itshape $\downarrow$261 &	\itshape $\downarrow$121 &	\itshape $\downarrow$1250 &	\itshape $\downarrow$317 &	\itshape $\downarrow$254 &	\itshape $\downarrow$151 &	\itshape $\downarrow$481 &	\itshape $\downarrow$287 &	\itshape $\downarrow$966 &	\itshape $\downarrow$213 &	\itshape $\downarrow$113 	\\	
BE	&	\itshape $\downarrow$68 &	\cellcolor{gray!80} &	\itshape $\downarrow$19 &	\itshape $\downarrow$119 &	\itshape $\downarrow$5 &	\itshape $\downarrow$1137 &	\itshape $\downarrow$184 &	\itshape $\downarrow$112 &	\itshape $\downarrow$34 &	\itshape $\downarrow$419 &	\itshape $\downarrow$153 &	\itshape $\downarrow$868 &	\itshape $\downarrow$67 &	\bfseries $\uparrow$13 	\\	
BG	&	\itshape $\downarrow$285 &	\itshape $\downarrow$375 &	\cellcolor{gray!80} &	\itshape $\downarrow$359 &	\itshape $\downarrow$227 &	\itshape $\downarrow$1380 &	\itshape $\downarrow$427 &	\itshape $\downarrow$354 &	\itshape $\downarrow$248 &	\itshape $\downarrow$490 &	\itshape $\downarrow$398 &	\itshape $\downarrow$1075 &	\itshape $\downarrow$319 &	\itshape $\downarrow$206 	\\	
CY	&	\itshape $\downarrow$1170 &	\itshape $\downarrow$1223 &	\itshape $\downarrow$1140 &	\cellcolor{gray!80} &	\itshape $\downarrow$1090 &	\itshape $\downarrow$2472 &	\itshape $\downarrow$1299 &	\itshape $\downarrow$1279 &	\itshape $\downarrow$1100 &	\itshape $\downarrow$929 &	\itshape $\downarrow$1278 &	\itshape $\downarrow$1852 &	\itshape $\downarrow$1227 &	\itshape $\downarrow$1073 	\\	
CZ	&	\itshape $\downarrow$105 &	\itshape $\downarrow$192 &	\itshape $\downarrow$35 &	\itshape $\downarrow$155 &	\cellcolor{gray!80} &	\itshape $\downarrow$1159 &	\itshape $\downarrow$211 &	\itshape $\downarrow$148 &	\itshape $\downarrow$54 &	\itshape $\downarrow$418 &	\itshape $\downarrow$180 &	\itshape $\downarrow$878 &	\itshape $\downarrow$109 &	\itshape $\downarrow$14 	\\	
DE	&	\bfseries $\uparrow$348 &	\bfseries $\uparrow$215 &	\bfseries $\uparrow$426 &	\bfseries $\uparrow$369 &	\bfseries $\uparrow$398 &	\cellcolor{gray!80} &	\bfseries $\uparrow$257 &	\bfseries $\uparrow$369 &	\bfseries $\uparrow$360 &	\bfseries $\uparrow$544 &	\bfseries $\uparrow$293 &	\bfseries $\uparrow$297 &	\bfseries $\uparrow$386 &	\bfseries $\uparrow$418 	\\	
DK	&	\itshape $\downarrow$434 &	\itshape $\downarrow$529 &	\itshape $\downarrow$392 &	\itshape $\downarrow$500 &	\itshape $\downarrow$374 &	\itshape $\downarrow$1545 &	\cellcolor{gray!80} &	\itshape $\downarrow$476 &	\itshape $\downarrow$398 &	\itshape $\downarrow$522 &	\itshape $\downarrow$541 &	\itshape $\downarrow$1178 &	\itshape $\downarrow$451 &	\itshape $\downarrow$344 	\\	
EE	&	\itshape $\downarrow$1067 &	\itshape $\downarrow$1138 &	\itshape $\downarrow$1042 &	\itshape $\downarrow$1192 &	\itshape $\downarrow$1002 &	\itshape $\downarrow$2349 &	\itshape $\downarrow$1206 &	\cellcolor{gray!80} &	\itshape $\downarrow$1013 &	\itshape $\downarrow$889 &	\itshape $\downarrow$1184 &	\itshape $\downarrow$1782 &	\itshape $\downarrow$1124 &	\itshape $\downarrow$963 	\\	
EL	&	\itshape $\downarrow$78 &	\itshape $\downarrow$185 &	\itshape $\downarrow$28 &	\itshape $\downarrow$126 &	\itshape $\downarrow$14 &	\itshape $\downarrow$1146 &	\itshape $\downarrow$192 &	\itshape $\downarrow$120 &	\cellcolor{gray!80} &	\itshape $\downarrow$403 &	\itshape $\downarrow$160 &	\itshape $\downarrow$877 &	\itshape $\downarrow$79 &	\bfseries $\uparrow$12 	\\	
ES	&	\bfseries $\uparrow$297 &	\bfseries $\uparrow$137 &	\bfseries $\uparrow$388 &	\bfseries $\uparrow$350 &	\bfseries $\uparrow$323 &	\bfseries $\uparrow$16 &	\bfseries $\uparrow$231 &	\bfseries $\uparrow$350 &	\bfseries $\uparrow$284 &	\cellcolor{gray!80} &	\bfseries $\uparrow$266 &	\bfseries $\uparrow$125 &	\bfseries $\uparrow$369 &	\bfseries $\uparrow$354 	\\	
FI	&	\itshape $\downarrow$454 &	\itshape $\downarrow$542 &	\itshape $\downarrow$408 &	\itshape $\downarrow$506 &	\itshape $\downarrow$385 &	\itshape $\downarrow$1555 &	\itshape $\downarrow$575 &	\itshape $\downarrow$499 &	\itshape $\downarrow$409 &	\itshape $\downarrow$527 &	\cellcolor{gray!80} &	\itshape $\downarrow$1182 &	\itshape $\downarrow$457 &	\itshape $\downarrow$358 	\\	
FR	&	\bfseries $\uparrow$346 &	\bfseries $\uparrow$207 &	\bfseries $\uparrow$429 &	\bfseries $\uparrow$382 &	\bfseries $\uparrow$393 &	\bfseries $\uparrow$160 &	\bfseries $\uparrow$266 &	\bfseries $\uparrow$380 &	\bfseries $\uparrow$355 &	\bfseries $\uparrow$526 &	\bfseries $\uparrow$302 &	\cellcolor{gray!80} &	\bfseries $\uparrow$400 &	\bfseries $\uparrow$416 	\\	
HR	&	\itshape $\downarrow$586 &	\itshape $\downarrow$682 &	\itshape $\downarrow$524 &	\itshape $\downarrow$649 &	\itshape $\downarrow$529 &	\itshape $\downarrow$1713 &	\itshape $\downarrow$695 &	\itshape $\downarrow$640 &	\itshape $\downarrow$553 &	\itshape $\downarrow$624 &	\itshape $\downarrow$667 &	\itshape $\downarrow$1294 &	\cellcolor{gray!80} &	\itshape $\downarrow$502 	\\	
HU	&	\itshape $\downarrow$124 &	\itshape $\downarrow$234 &	\itshape $\downarrow$62 &	\itshape $\downarrow$186 &	\itshape $\downarrow$68 &	\itshape $\downarrow$1191 &	\itshape $\downarrow$253 &	\itshape $\downarrow$171 &	\itshape $\downarrow$96 &	\itshape $\downarrow$425 &	\itshape $\downarrow$222 &	\itshape $\downarrow$887 &	\itshape $\downarrow$139 &	\cellcolor{gray!80}	\\	
IT	&	\bfseries $\uparrow$327 &	\bfseries $\uparrow$184 &	\bfseries $\uparrow$406 &	\bfseries $\uparrow$367 &	\bfseries $\uparrow$368 &	\bfseries $\uparrow$54 &	\bfseries $\uparrow$241 &	\bfseries $\uparrow$367 &	\bfseries $\uparrow$330 &	\bfseries $\uparrow$535 &	\bfseries $\uparrow$277 &	\bfseries $\uparrow$71 &	\bfseries $\uparrow$375 &	\bfseries $\uparrow$393 	\\	
IE	&	\itshape $\downarrow$552 &	\itshape $\downarrow$637 &	\itshape $\downarrow$484 &	\itshape $\downarrow$604 &	\itshape $\downarrow$485 &	\itshape $\downarrow$1655 &	\itshape $\downarrow$662 &	\itshape $\downarrow$600 &	\itshape $\downarrow$513 &	\itshape $\downarrow$590 &	\itshape $\downarrow$634 &	\itshape $\downarrow$1269 &	\itshape $\downarrow$540 &	\itshape $\downarrow$469 	\\	
LT	&	\itshape $\downarrow$739 &	\itshape $\downarrow$810 &	\itshape $\downarrow$678 &	\itshape $\downarrow$848 &	\itshape $\downarrow$672 &	\itshape $\downarrow$1914 &	\itshape $\downarrow$855 &	\itshape $\downarrow$841 &	\itshape $\downarrow$685 &	\itshape $\downarrow$769 &	\itshape $\downarrow$830 &	\itshape $\downarrow$1445 &	\itshape $\downarrow$768 &	\itshape $\downarrow$649 	\\	
LU	&	\itshape $\downarrow$1248 &	\itshape $\downarrow$1300 &	\itshape $\downarrow$1206 &	\itshape $\downarrow$1401 &	\itshape $\downarrow$1176 &	\itshape $\downarrow$2568 &	\itshape $\downarrow$1389 &	\itshape $\downarrow$1369 &	\itshape $\downarrow$1168 &	\itshape $\downarrow$962 &	\itshape $\downarrow$1364 &	\itshape $\downarrow$1931 &	\itshape $\downarrow$1311 &	\itshape $\downarrow$1136 	\\	
LV	&	\itshape $\downarrow$901 &	\itshape $\downarrow$975 &	\itshape $\downarrow$865 &	\itshape $\downarrow$1028 &	\itshape $\downarrow$826 &	\itshape $\downarrow$2130 &	\itshape $\downarrow$1045 &	\itshape $\downarrow$1014 &	\itshape $\downarrow$850 &	\itshape $\downarrow$841 &	\itshape $\downarrow$1018 &	\itshape $\downarrow$1629 &	\itshape $\downarrow$960 &	\itshape $\downarrow$809 	\\	
MT	&	\itshape $\downarrow$1308 &	\itshape $\downarrow$1368 &	\itshape $\downarrow$1278 &	\itshape $\downarrow$1451 &	\itshape $\downarrow$1234 &	\itshape $\downarrow$2632 &	\itshape $\downarrow$1444 &	\itshape $\downarrow$1437 &	\itshape $\downarrow$1247 &	\itshape $\downarrow$996 &	\itshape $\downarrow$1423 &	\itshape $\downarrow$1986 &	\itshape $\downarrow$1359 &	\itshape $\downarrow$1208 	\\	
NL	&	\bfseries $\uparrow$96 &	\bfseries $\uparrow$6 &	\bfseries $\uparrow$141 &	\bfseries $\uparrow$63 &	\bfseries $\uparrow$172 &	\itshape $\downarrow$1008 &	\itshape $\downarrow$31 &	\bfseries $\uparrow$66 &	\bfseries $\uparrow$142 &	\itshape $\downarrow$455 &	\bfseries $\uparrow$1 &	\itshape $\downarrow$853 &	\bfseries $\uparrow$83 &	\bfseries $\uparrow$187 	\\	
PL	&	\bfseries $\uparrow$213 &	\bfseries $\uparrow$62 &	\bfseries $\uparrow$292 &	\bfseries $\uparrow$235 &	\bfseries $\uparrow$244 &	\itshape $\downarrow$645 &	\bfseries $\uparrow$134 &	\bfseries $\uparrow$237 &	\bfseries $\uparrow$209 &	\itshape $\downarrow$873 &	\bfseries $\uparrow$167 &	\itshape $\downarrow$582 &	\bfseries $\uparrow$263 &	\bfseries $\uparrow$277 	\\	
PT	&	\itshape $\downarrow$112 &	\itshape $\downarrow$202 &	\itshape $\downarrow$42 &	\itshape $\downarrow$160 &	\itshape $\downarrow$28 &	\itshape $\downarrow$1163 &	\itshape $\downarrow$216 &	\itshape $\downarrow$153 &	\itshape $\downarrow$60 &	\itshape $\downarrow$420 &	\itshape $\downarrow$184 &	\itshape $\downarrow$879 &	\itshape $\downarrow$118 &	\itshape $\downarrow$26 	\\	
RO	&	\bfseries $\uparrow$186 &	\bfseries $\uparrow$82 &	\bfseries $\uparrow$253 &	\bfseries $\uparrow$134 &	\bfseries $\uparrow$251 &	\itshape $\downarrow$1043 &	\bfseries $\uparrow$69 &	\bfseries $\uparrow$136 &	\bfseries $\uparrow$219 &	\itshape $\downarrow$582 &	\bfseries $\uparrow$101 &	\itshape $\downarrow$889 &	\bfseries $\uparrow$187 &	\bfseries $\uparrow$264 	\\	
SE	&	\itshape $\downarrow$133 &	\itshape $\downarrow$239 &	\itshape $\downarrow$79 &	\itshape $\downarrow$195 &	\itshape $\downarrow$76 &	\itshape $\downarrow$1196 &	\itshape $\downarrow$260 &	\itshape $\downarrow$188 &	\itshape $\downarrow$103 &	\itshape $\downarrow$418 &	\itshape $\downarrow$229 &	\itshape $\downarrow$885 &	\itshape $\downarrow$144 &	\itshape $\downarrow$50 	\\	
SI	&	\itshape $\downarrow$868 &	\itshape $\downarrow$934 &	\itshape $\downarrow$843 &	\itshape $\downarrow$1001 &	\itshape $\downarrow$801 &	\itshape $\downarrow$2102 &	\itshape $\downarrow$1013 &	\itshape $\downarrow$985 &	\itshape $\downarrow$827 &	\itshape $\downarrow$829 &	\itshape $\downarrow$990 &	\itshape $\downarrow$1611 &	\itshape $\downarrow$932 &	\itshape $\downarrow$777 	\\	
SK	&	\itshape $\downarrow$465 &	\itshape $\downarrow$553 &	\itshape $\downarrow$416 &	\itshape $\downarrow$513 &	\itshape $\downarrow$392 &	\itshape $\downarrow$1565 &	\itshape $\downarrow$582 &	\itshape $\downarrow$506 &	\itshape $\downarrow$421 &	\itshape $\downarrow$526 &	\itshape $\downarrow$553 &	\itshape $\downarrow$1187 &	\itshape $\downarrow$465 &	\itshape $\downarrow$366 	\\	
UK	&	\bfseries $\uparrow$346 &	\bfseries $\uparrow$201 &	\bfseries $\uparrow$425 &	\bfseries $\uparrow$376 &	\bfseries $\uparrow$387 &	\bfseries $\uparrow$136 &	\bfseries $\uparrow$262 &	\bfseries $\uparrow$376 &	\bfseries $\uparrow$347 &	\bfseries $\uparrow$541 &	\bfseries $\uparrow$298 &	\bfseries $\uparrow$142 &	\bfseries $\uparrow$398 &	\bfseries $\uparrow$415 	\\	
 \end{tabularx}%
}
\caption{The impact of any member state leaving the 28-member European Union}
  \label{bazispontos tablazat 1}%
\end{footnotesize}
\end{table}

\addtocounter{table}{-1}
\begin{table}
\centering
\begin{footnotesize}
\centerline{
\rowcolors{1}{gray!20}{}
\begin{tabularx}{1.15\textwidth}{l CCCCC CCCCC CCCC}  \toprule \hiderowcolors
	& IT & IE & LT & LU & LV & MT & NL & PL & PT & RO & SE &SI & SK &UK \\ \hline
\showrowcolors
\hline
AT	&	\itshape $\downarrow$652 &	\itshape $\downarrow$276 &	\itshape $\downarrow$233 &	\itshape $\downarrow$275 &	\itshape $\downarrow$258 &	\itshape $\downarrow$264 &	\itshape $\downarrow$359 &	\bfseries $\uparrow$215 &	\itshape $\downarrow$163 &	\bfseries $\uparrow$153 &	\itshape $\downarrow$349 &	\itshape $\downarrow$261 &	\itshape $\downarrow$203 &	\itshape $\downarrow$252 	\\	
BE	&	\itshape $\downarrow$502 &	\itshape $\downarrow$140 &	\itshape $\downarrow$88 &	\itshape $\downarrow$129 &	\itshape $\downarrow$107 &	\itshape $\downarrow$119 &	\itshape $\downarrow$227 &	\bfseries $\uparrow$243 &	\itshape $\downarrow$31 &	\bfseries $\uparrow$272 &	\itshape $\downarrow$221 &	\itshape $\downarrow$112 &	\itshape $\downarrow$65 &	\itshape $\downarrow$219 	\\	
BG	&	\itshape $\downarrow$705 &	\itshape $\downarrow$386 &	\itshape $\downarrow$353 &	\itshape $\downarrow$371 &	\itshape $\downarrow$365 &	\itshape $\downarrow$361 &	\itshape $\downarrow$445 &	\bfseries $\uparrow$164 &	\itshape $\downarrow$251 &	\bfseries $\uparrow$42 &	\itshape $\downarrow$436 &	\itshape $\downarrow$368 &	\itshape $\downarrow$315 &	\itshape $\downarrow$407 	\\	
CY	&	\itshape $\downarrow$1400 &	\itshape $\downarrow$1287 &	\itshape $\downarrow$1265 &	\itshape $\downarrow$1328 &	\itshape $\downarrow$1285 &	\itshape $\downarrow$1320 &	\itshape $\downarrow$1280 &	\itshape $\downarrow$495 &	\itshape $\downarrow$1099 &	\itshape $\downarrow$890 &	\itshape $\downarrow$1292 &	\itshape $\downarrow$1282 &	\itshape $\downarrow$1208 &	\itshape $\downarrow$1234 	\\	
CZ	&	\itshape $\downarrow$547 &	\itshape $\downarrow$181 &	\itshape $\downarrow$129 &	\itshape $\downarrow$171 &	\itshape $\downarrow$139 &	\itshape $\downarrow$161 &	\itshape $\downarrow$257 &	\bfseries $\uparrow$248 &	\itshape $\downarrow$50 &	\bfseries $\uparrow$247 &	\itshape $\downarrow$248 &	\itshape $\downarrow$141 &	\itshape $\downarrow$93 &	\itshape $\downarrow$238 	\\	
DE	&	\bfseries $\uparrow$586 &	\bfseries $\uparrow$309 &	\bfseries $\uparrow$378 &	\bfseries $\uparrow$360 &	\bfseries $\uparrow$373 &	\bfseries $\uparrow$370 &	\bfseries $\uparrow$90 &	\bfseries $\uparrow$826 &	\bfseries $\uparrow$367 &	\bfseries $\uparrow$582 &	\bfseries $\uparrow$178 &	\bfseries $\uparrow$366 &	\bfseries $\uparrow$387 &	\bfseries $\uparrow$1002 	\\	
DK	&	\itshape $\downarrow$822 &	\itshape $\downarrow$523 &	\itshape $\downarrow$473 &	\itshape $\downarrow$513 &	\itshape $\downarrow$485 &	\itshape $\downarrow$503 &	\itshape $\downarrow$577 &	\bfseries $\uparrow$11 &	\itshape $\downarrow$400 &	\itshape $\downarrow$90 &	\itshape $\downarrow$576 &	\itshape $\downarrow$487 &	\itshape $\downarrow$459 &	\itshape $\downarrow$535 	\\	
EE	&	\itshape $\downarrow$1322 &	\itshape $\downarrow$1187 &	\itshape $\downarrow$1168 &	\itshape $\downarrow$1206 &	\itshape $\downarrow$1186 &	\itshape $\downarrow$1202 &	\itshape $\downarrow$1195 &	\itshape $\downarrow$437 &	\itshape $\downarrow$1012 &	\itshape $\downarrow$790 &	\itshape $\downarrow$1187 &	\itshape $\downarrow$1184 &	\itshape $\downarrow$1112 &	\itshape $\downarrow$1132 	\\	
EL	&	\itshape $\downarrow$545 &	\itshape $\downarrow$151 &	\itshape $\downarrow$110 &	\itshape $\downarrow$139 &	\itshape $\downarrow$119 &	\itshape $\downarrow$129 &	\itshape $\downarrow$236 &	\bfseries $\uparrow$246 &	\itshape $\downarrow$39 &	\bfseries $\uparrow$260 &	\itshape $\downarrow$223 &	\itshape $\downarrow$123 &	\itshape $\downarrow$72 &	\itshape $\downarrow$228 	\\	
ES	&	\bfseries $\uparrow$502 &	\bfseries $\uparrow$287 &	\bfseries $\uparrow$367 &	\bfseries $\uparrow$340 &	\bfseries $\uparrow$366 &	\bfseries $\uparrow$352 &	\itshape $\downarrow$79 &	\itshape $\downarrow$48 &	\bfseries $\uparrow$295 &	\bfseries $\uparrow$318 &	\bfseries $\uparrow$116 &	\bfseries $\uparrow$359 &	\bfseries $\uparrow$359 &	\bfseries $\uparrow$861 	\\	
FI	&	\itshape $\downarrow$825 &	\itshape $\downarrow$532 &	\itshape $\downarrow$480 &	\itshape $\downarrow$520 &	\itshape $\downarrow$494 &	\itshape $\downarrow$515 &	\itshape $\downarrow$580 &	\bfseries $\uparrow$3 &	\itshape $\downarrow$407 &	\itshape $\downarrow$102 &	\itshape $\downarrow$588 &	\itshape $\downarrow$495 &	\itshape $\downarrow$464 &	\itshape $\downarrow$536 	\\	
FR	&	\bfseries $\uparrow$474 &	\bfseries $\uparrow$323 &	\bfseries $\uparrow$394 &	\bfseries $\uparrow$372 &	\bfseries $\uparrow$393 &	\bfseries $\uparrow$382 &	\bfseries $\uparrow$47 &	\bfseries $\uparrow$732 &	\bfseries $\uparrow$364 &	\bfseries $\uparrow$487 &	\bfseries $\uparrow$176 &	\bfseries $\uparrow$386 &	\bfseries $\uparrow$395 &	\bfseries $\uparrow$853 	\\	
HR	&	\itshape $\downarrow$938 &	\itshape $\downarrow$661 &	\itshape $\downarrow$618 &	\itshape $\downarrow$662 &	\itshape $\downarrow$643 &	\itshape $\downarrow$669 &	\itshape $\downarrow$742 &	\itshape $\downarrow$86 &	\itshape $\downarrow$548 &	\itshape $\downarrow$263 &	\itshape $\downarrow$731 &	\itshape $\downarrow$643 &	\itshape $\downarrow$588 &	\itshape $\downarrow$638 	\\	
HU	&	\itshape $\downarrow$553 &	\itshape $\downarrow$208 &	\itshape $\downarrow$163 &	\itshape $\downarrow$205 &	\itshape $\downarrow$179 &	\itshape $\downarrow$194 &	\itshape $\downarrow$286 &	\bfseries $\uparrow$233 &	\itshape $\downarrow$97 &	\bfseries $\uparrow$217 &	\itshape $\downarrow$279 &	\itshape $\downarrow$184 &	\itshape $\downarrow$136 &	\itshape $\downarrow$239 	\\	
IT	&	\cellcolor{gray!80} &	\bfseries $\uparrow$294 &	\bfseries $\uparrow$370 &	\bfseries $\uparrow$358 &	\bfseries $\uparrow$370 &	\bfseries $\uparrow$369 &	\bfseries $\uparrow$21 &	\bfseries $\uparrow$647 &	\bfseries $\uparrow$338 &	\bfseries $\uparrow$460 &	\bfseries $\uparrow$154 &	\bfseries $\uparrow$363 &	\bfseries $\uparrow$370 &	\bfseries $\uparrow$791 	\\	
IE	&	\itshape $\downarrow$907 &	\cellcolor{gray!80} &	\itshape $\downarrow$569 &	\itshape $\downarrow$619 &	\itshape $\downarrow$595 &	\itshape $\downarrow$606 &	\itshape $\downarrow$694 &	\itshape $\downarrow$57 &	\itshape $\downarrow$506 &	\itshape $\downarrow$219 &	\itshape $\downarrow$697 &	\itshape $\downarrow$595 &	\itshape $\downarrow$554 &	\itshape $\downarrow$612 	\\	
LT	&	\itshape $\downarrow$1078 &	\itshape $\downarrow$820 &	\cellcolor{gray!80} &	\itshape $\downarrow$861 &	\itshape $\downarrow$848 &	\itshape $\downarrow$861 &	\itshape $\downarrow$912 &	\itshape $\downarrow$157 &	\itshape $\downarrow$689 &	\itshape $\downarrow$431 &	\itshape $\downarrow$876 &	\itshape $\downarrow$847 &	\itshape $\downarrow$751 &	\itshape $\downarrow$792 	\\	
LU	&	\itshape $\downarrow$1482 &	\itshape $\downarrow$1375 &	\itshape $\downarrow$1353 &	\cellcolor{gray!80} &	\itshape $\downarrow$1374 &	\itshape $\downarrow$1407 &	\itshape $\downarrow$1364 &	\itshape $\downarrow$557 &	\itshape $\downarrow$1197 &	\itshape $\downarrow$961 &	\itshape $\downarrow$1358 &	\itshape $\downarrow$1369 &	\itshape $\downarrow$1296 &	\itshape $\downarrow$1335 	\\	
LV	&	\itshape $\downarrow$1198 &	\itshape $\downarrow$1021 &	\itshape $\downarrow$1004 &	\itshape $\downarrow$1053 &	\cellcolor{gray!80} &	\itshape $\downarrow$1046 &	\itshape $\downarrow$1072 &	\itshape $\downarrow$294 &	\itshape $\downarrow$846 &	\itshape $\downarrow$613 &	\itshape $\downarrow$1032 &	\itshape $\downarrow$1020 &	\itshape $\downarrow$943 &	\itshape $\downarrow$1004 	\\	
MT	&	\itshape $\downarrow$1517 &	\itshape $\downarrow$1425 &	\itshape $\downarrow$1400 &	\itshape $\downarrow$1490 &	\itshape $\downarrow$1446 &	\cellcolor{gray!80} &	\itshape $\downarrow$1419 &	\itshape $\downarrow$631 &	\itshape $\downarrow$1248 &	\itshape $\downarrow$1015 &	\itshape $\downarrow$1429 &	\itshape $\downarrow$1439 &	\itshape $\downarrow$1357 &	\itshape $\downarrow$1366 	\\	
NL	&	\itshape $\downarrow$567 &	\bfseries $\uparrow$12 &	\bfseries $\uparrow$67 &	\bfseries $\uparrow$52 &	\bfseries $\uparrow$67 &	\bfseries $\uparrow$58 &	\cellcolor{gray!80} &	\bfseries $\uparrow$223 &	\bfseries $\uparrow$143 &	\bfseries $\uparrow$503 &	\itshape $\downarrow$50 &	\bfseries $\uparrow$61 &	\bfseries $\uparrow$90 &	\itshape $\downarrow$200 	\\	
PL	&	\itshape $\downarrow$287 &	\bfseries $\uparrow$181 &	\bfseries $\uparrow$244 &	\bfseries $\uparrow$226 &	\bfseries $\uparrow$234 &	\bfseries $\uparrow$238 &	\itshape $\downarrow$155 &	\cellcolor{gray!80} &	\bfseries $\uparrow$216 &	\bfseries $\uparrow$234 &	\bfseries $\uparrow$36 &	\bfseries $\uparrow$230 &	\bfseries $\uparrow$257 &	\bfseries $\uparrow$91 	\\	
PT	&	\itshape $\downarrow$548 &	\itshape $\downarrow$185 &	\itshape $\downarrow$134 &	\itshape $\downarrow$177 &	\itshape $\downarrow$144 &	\itshape $\downarrow$166 &	\itshape $\downarrow$268 &	\bfseries $\uparrow$238 &	\cellcolor{gray!80} &	\bfseries $\uparrow$237 &	\itshape $\downarrow$261 &	\itshape $\downarrow$147 &	\itshape $\downarrow$96 &	\itshape $\downarrow$236 	\\	
RO	&	\itshape $\downarrow$634 &	\bfseries $\uparrow$111 &	\bfseries $\uparrow$163 &	\bfseries $\uparrow$123 &	\bfseries $\uparrow$146 &	\bfseries $\uparrow$133 &	\bfseries $\uparrow$55 &	\bfseries $\uparrow$154 &	\bfseries $\uparrow$220 &	\cellcolor{gray!80} &	\bfseries $\uparrow$23 &	\bfseries $\uparrow$140 &	\bfseries $\uparrow$191 &	\itshape $\downarrow$250 	\\	
SE	&	\itshape $\downarrow$553 &	\itshape $\downarrow$218 &	\itshape $\downarrow$171 &	\itshape $\downarrow$206 &	\itshape $\downarrow$184 &	\itshape $\downarrow$198 &	\itshape $\downarrow$299 &	\bfseries $\uparrow$226 &	\itshape $\downarrow$101 &	\bfseries $\uparrow$206 &	\cellcolor{gray!80} &	\itshape $\downarrow$188 &	\itshape $\downarrow$143 &	\itshape $\downarrow$232 	\\	
SI	&	\itshape $\downarrow$1178 &	\itshape $\downarrow$996 &	\itshape $\downarrow$970 &	\itshape $\downarrow$1029 &	\itshape $\downarrow$997 &	\itshape $\downarrow$1019 &	\itshape $\downarrow$1053 &	\itshape $\downarrow$270 &	\itshape $\downarrow$821 &	\itshape $\downarrow$587 &	\itshape $\downarrow$1002 &	\cellcolor{gray!80} &	\itshape $\downarrow$913 &	\itshape $\downarrow$982 	\\	
SK	&	\itshape $\downarrow$849 &	\itshape $\downarrow$539 &	\itshape $\downarrow$492 &	\itshape $\downarrow$529 &	\itshape $\downarrow$498 &	\itshape $\downarrow$521 &	\itshape $\downarrow$595 &	\itshape $\downarrow$0 &	\itshape $\downarrow$412 &	\itshape $\downarrow$109 &	\itshape $\downarrow$596 &	\itshape $\downarrow$499 &	\cellcolor{gray!80} &	\itshape $\downarrow$549 	\\	
UK	&	\bfseries $\uparrow$434 &	\bfseries $\uparrow$317 &	\bfseries $\uparrow$392 &	\bfseries $\uparrow$366 &	\bfseries $\uparrow$391 &	\bfseries $\uparrow$378 &	\bfseries $\uparrow$32 &	\bfseries $\uparrow$722 &	\bfseries $\uparrow$358 &	\bfseries $\uparrow$484 &	\bfseries $\uparrow$176 &	\bfseries $\uparrow$384 &	\bfseries $\uparrow$391 &	\cellcolor{gray!80}	\\	
    \end{tabularx}%
    }
\caption{ The impact of any member state leaving the 28-member European Union (continued)}    
  \label{bazispontos tablazat 1b}%
\end{footnotesize}
\end{table}%

%% file: Appendix_C.tex
\begin{table}[!htbp]
  \centering
\begin{footnotesize}
\centerline{
\rowcolors{1}{gray!20}{}
    \begin{tabularx}{1.15\textwidth}{l CCCCC CCCCC CCCC} \toprule \hiderowcolors
	 & AT & BE & BG & CY & CZ & DE & DK & EE & EL & ES & FI & FR & HR & HU \\ \hline
\showrowcolors
	AT &\cellcolor{gray!80} &\bfseries $\uparrow$337 &\bfseries $\uparrow$599 &\bfseries $\uparrow$553 &\bfseries $\uparrow$553 & \itshape $\downarrow$1038 &\bfseries $\uparrow$418 &\bfseries $\uparrow$553 &\bfseries $\uparrow$502 & \itshape $\downarrow$38 &\bfseries $\uparrow$451 & \itshape $\downarrow$485 &\bfseries $\uparrow$557 &\bfseries $\uparrow$570 \\ 
	BE &\bfseries $\uparrow$356 &\cellcolor{gray!80} &\bfseries $\uparrow$482 &\bfseries $\uparrow$365 &\bfseries $\uparrow$389 & \itshape $\downarrow$1059 &\bfseries $\uparrow$253 &\bfseries $\uparrow$377 &\bfseries $\uparrow$329 & \itshape $\downarrow$172 &\bfseries $\uparrow$287 & \itshape $\downarrow$618 &\bfseries $\uparrow$390 &\bfseries $\uparrow$415 \\ 
	BG &\bfseries $\uparrow$748 &\bfseries $\uparrow$578 &\cellcolor{gray!80} &\bfseries $\uparrow$711 &\bfseries $\uparrow$811 & \itshape $\downarrow$884 &\bfseries $\uparrow$708 &\bfseries $\uparrow$717 &\bfseries $\uparrow$745 &\bfseries $\uparrow$184 &\bfseries $\uparrow$742 & \itshape $\downarrow$279 &\bfseries $\uparrow$855 &\bfseries $\uparrow$816 \\ 
	CY &\bfseries $\uparrow$2427 &\bfseries $\uparrow$2238 &\bfseries $\uparrow$2569 &\cellcolor{gray!80} &\bfseries $\uparrow$2548 & \itshape $\downarrow$39 &\bfseries $\uparrow$2270 &\bfseries $\uparrow$2352 &\bfseries $\uparrow$2434 &\bfseries $\uparrow$2003 &\bfseries $\uparrow$2322 &\bfseries $\uparrow$1159 &\bfseries $\uparrow$2379 &\bfseries $\uparrow$2535 \\ 
	CZ &\bfseries $\uparrow$370 &\bfseries $\uparrow$244 &\bfseries $\uparrow$488 &\bfseries $\uparrow$408 &\cellcolor{gray!80} & \itshape $\downarrow$1043 &\bfseries $\uparrow$305 &\bfseries $\uparrow$417 &\bfseries $\uparrow$409 & \itshape $\downarrow$129 &\bfseries $\uparrow$338 & \itshape $\downarrow$581 &\bfseries $\uparrow$439 &\bfseries $\uparrow$487 \\ 
	DE & \itshape $\downarrow$343 & \itshape $\downarrow$462 & \itshape $\downarrow$298 & \itshape $\downarrow$364 & \itshape $\downarrow$290 &\cellcolor{gray!80} & \itshape $\downarrow$473 & \itshape $\downarrow$363 & \itshape $\downarrow$311 & \itshape $\downarrow$346 & \itshape $\downarrow$436 & \itshape $\downarrow$402 & \itshape $\downarrow$335 & \itshape $\downarrow$255 \\ 
	DK &\bfseries $\uparrow$1052 &\bfseries $\uparrow$885 &\bfseries $\uparrow$1067 &\bfseries $\uparrow$926 &\bfseries $\uparrow$1105 & \itshape $\downarrow$763 &\cellcolor{gray!80} &\bfseries $\uparrow$930 &\bfseries $\uparrow$1034 &\bfseries $\uparrow$420 &\bfseries $\uparrow$907 & \itshape $\downarrow$17 &\bfseries $\uparrow$985 &\bfseries $\uparrow$1121 \\ 
	EE &\bfseries $\uparrow$2207 &\bfseries $\uparrow$1990 &\bfseries $\uparrow$2349 &\bfseries $\uparrow$2069 &\bfseries $\uparrow$2249 & \itshape $\downarrow$177 &\bfseries $\uparrow$2062 &\cellcolor{gray!80} &\bfseries $\uparrow$2171 &\bfseries $\uparrow$1753 &\bfseries $\uparrow$2094 &\bfseries $\uparrow$924 &\bfseries $\uparrow$2171 &\bfseries $\uparrow$2248 \\ 
	EL &\bfseries $\uparrow$380 &\bfseries $\uparrow$188 &\bfseries $\uparrow$447 &\bfseries $\uparrow$384 &\bfseries $\uparrow$414 & \itshape $\downarrow$1056 &\bfseries $\uparrow$260 &\bfseries $\uparrow$387 &\cellcolor{gray!80} & \itshape $\downarrow$167 &\bfseries $\uparrow$297 & \itshape $\downarrow$604 &\bfseries $\uparrow$403 &\bfseries $\uparrow$438 \\ 
	ES & \itshape $\downarrow$358 & \itshape $\downarrow$503 & \itshape $\downarrow$271 & \itshape $\downarrow$319 & \itshape $\downarrow$319 & \itshape $\downarrow$495 & \itshape $\downarrow$442 & \itshape $\downarrow$322 & \itshape $\downarrow$354 &\cellcolor{gray!80} & \itshape $\downarrow$405 & \itshape $\downarrow$617 & \itshape $\downarrow$299 & \itshape $\downarrow$287 \\ 
	FI &\bfseries $\uparrow$1077 &\bfseries $\uparrow$891 &\bfseries $\uparrow$1099 &\bfseries $\uparrow$936 &\bfseries $\uparrow$1133 & \itshape $\downarrow$746 &\bfseries $\uparrow$888 &\bfseries $\uparrow$958 &\bfseries $\uparrow$1066 &\bfseries $\uparrow$444 &\cellcolor{gray!80} &\bfseries $\uparrow$10 &\bfseries $\uparrow$999 &\bfseries $\uparrow$1135 \\ 
	FR & \itshape $\downarrow$373 & \itshape $\downarrow$511 & \itshape $\downarrow$320 & \itshape $\downarrow$364 & \itshape $\downarrow$334 & \itshape $\downarrow$249 & \itshape $\downarrow$485 & \itshape $\downarrow$366 & \itshape $\downarrow$359 & \itshape $\downarrow$530 & \itshape $\downarrow$448 &\cellcolor{gray!80} & \itshape $\downarrow$342 & \itshape $\downarrow$303 \\ 
	HR &\bfseries $\uparrow$1173 &\bfseries $\uparrow$1044 &\bfseries $\uparrow$1302 &\bfseries $\uparrow$1175 &\bfseries $\uparrow$1263 & \itshape $\downarrow$635 &\bfseries $\uparrow$1063 &\bfseries $\uparrow$1180 &\bfseries $\uparrow$1204 &\bfseries $\uparrow$714 &\bfseries $\uparrow$1097 &\bfseries $\uparrow$132 &\cellcolor{gray!80} &\bfseries $\uparrow$1265 \\ 
	HU &\bfseries $\uparrow$403 &\bfseries $\uparrow$292 &\bfseries $\uparrow$531 &\bfseries $\uparrow$451 &\bfseries $\uparrow$455 & \itshape $\downarrow$1031 &\bfseries $\uparrow$345 &\bfseries $\uparrow$450 &\bfseries $\uparrow$454 & \itshape $\downarrow$102 &\bfseries $\uparrow$378 & \itshape $\downarrow$563 &\bfseries $\uparrow$467 &\cellcolor{gray!80} \\ 
	IT & \itshape $\downarrow$354 & \itshape $\downarrow$473 & \itshape $\downarrow$287 & \itshape $\downarrow$359 & \itshape $\downarrow$300 & \itshape $\downarrow$383 & \itshape $\downarrow$469 & \itshape $\downarrow$359 & \itshape $\downarrow$324 & \itshape $\downarrow$316 & \itshape $\downarrow$432 & \itshape $\downarrow$980 & \itshape $\downarrow$333 & \itshape $\downarrow$262 \\ 
	IE &\bfseries $\uparrow$1109 &\bfseries $\uparrow$1081 &\bfseries $\uparrow$1224 &\bfseries $\uparrow$1076 &\bfseries $\uparrow$1209 & \itshape $\downarrow$673 &\bfseries $\uparrow$1027 &\bfseries $\uparrow$1080 &\bfseries $\uparrow$1259 &\bfseries $\uparrow$627 &\bfseries $\uparrow$1060 &\bfseries $\uparrow$93 &\bfseries $\uparrow$1108 &\bfseries $\uparrow$1206 \\ 
	LT &\bfseries $\uparrow$1494 &\bfseries $\uparrow$1353 &\bfseries $\uparrow$1622 &\bfseries $\uparrow$1522 &\bfseries $\uparrow$1611 & \itshape $\downarrow$485 &\bfseries $\uparrow$1417 &\bfseries $\uparrow$1545 &\bfseries $\uparrow$1526 &\bfseries $\uparrow$1093 &\bfseries $\uparrow$1451 &\bfseries $\uparrow$414 &\bfseries $\uparrow$1561 &\bfseries $\uparrow$1603 \\ 
	LU &\bfseries $\uparrow$2561 &\bfseries $\uparrow$2456 &\bfseries $\uparrow$2736 &\bfseries $\uparrow$2483 &\bfseries $\uparrow$2769 &\bfseries $\uparrow$51 &\bfseries $\uparrow$2435 &\bfseries $\uparrow$2498 &\bfseries $\uparrow$2645 &\bfseries $\uparrow$2203 &\bfseries $\uparrow$2465 &\bfseries $\uparrow$1313 &\bfseries $\uparrow$2549 &\bfseries $\uparrow$2711 \\ 
	LV &\bfseries $\uparrow$1866 &\bfseries $\uparrow$1706 &\bfseries $\uparrow$2051 &\bfseries $\uparrow$1790 &\bfseries $\uparrow$1960 & \itshape $\downarrow$314 &\bfseries $\uparrow$1804 &\bfseries $\uparrow$1822 &\bfseries $\uparrow$1877 &\bfseries $\uparrow$1475 &\bfseries $\uparrow$1843 &\bfseries $\uparrow$704 &\bfseries $\uparrow$1923 &\bfseries $\uparrow$1972 \\ 
	MT &\bfseries $\uparrow$2693 &\bfseries $\uparrow$2575 &\bfseries $\uparrow$2836 &\bfseries $\uparrow$2553 &\bfseries $\uparrow$2823 &\bfseries $\uparrow$102 &\bfseries $\uparrow$2529 &\bfseries $\uparrow$2580 &\bfseries $\uparrow$2773 &\bfseries $\uparrow$2331 &\bfseries $\uparrow$2561 &\bfseries $\uparrow$1429 &\bfseries $\uparrow$2660 &\bfseries $\uparrow$2795 \\ 
	NL &\bfseries $\uparrow$54 & \itshape $\downarrow$99 &\bfseries $\uparrow$166 &\bfseries $\uparrow$116 &\bfseries $\uparrow$96 & \itshape $\downarrow$1147 &\bfseries $\uparrow$10 &\bfseries $\uparrow$117 &\bfseries $\uparrow$54 & \itshape $\downarrow$396 &\bfseries $\uparrow$45 & \itshape $\downarrow$787 &\bfseries $\uparrow$148 &\bfseries $\uparrow$123 \\ 
	PL & \itshape $\downarrow$468 & \itshape $\downarrow$638 & \itshape $\downarrow$372 & \itshape $\downarrow$365 & \itshape $\downarrow$444 &\bfseries $\uparrow$201 & \itshape $\downarrow$528 & \itshape $\downarrow$368 & \itshape $\downarrow$490 &\bfseries $\uparrow$1535 & \itshape $\downarrow$490 &\bfseries $\uparrow$196 & \itshape $\downarrow$376 & \itshape $\downarrow$404 \\ 
	PT &\bfseries $\uparrow$381 &\bfseries $\uparrow$257 &\bfseries $\uparrow$494 &\bfseries $\uparrow$409 &\bfseries $\uparrow$474 & \itshape $\downarrow$1040 &\bfseries $\uparrow$305 &\bfseries $\uparrow$422 &\bfseries $\uparrow$417 & \itshape $\downarrow$119 &\bfseries $\uparrow$341 & \itshape $\downarrow$575 &\bfseries $\uparrow$447 &\bfseries $\uparrow$497 \\ 
	RO & \itshape $\downarrow$45 & \itshape $\downarrow$204 &\bfseries $\uparrow$40 &\bfseries $\uparrow$7 & \itshape $\downarrow$6 & \itshape $\downarrow$1180 & \itshape $\downarrow$123 &\bfseries $\uparrow$9 & \itshape $\downarrow$50 & \itshape $\downarrow$421 & \itshape $\downarrow$89 & \itshape $\downarrow$852 &\bfseries $\uparrow$19 &\bfseries $\uparrow$17 \\ 
	SE &\bfseries $\uparrow$417 &\bfseries $\uparrow$249 &\bfseries $\uparrow$529 &\bfseries $\uparrow$463 &\bfseries $\uparrow$472 & \itshape $\downarrow$1037 &\bfseries $\uparrow$342 &\bfseries $\uparrow$463 &\bfseries $\uparrow$408 & \itshape $\downarrow$102 &\bfseries $\uparrow$380 & \itshape $\downarrow$552 &\bfseries $\uparrow$461 &\bfseries $\uparrow$488 \\ 
	SI &\bfseries $\uparrow$1825 &\bfseries $\uparrow$1657 &\bfseries $\uparrow$2002 &\bfseries $\uparrow$1773 &\bfseries $\uparrow$1920 & \itshape $\downarrow$333 &\bfseries $\uparrow$1783 &\bfseries $\uparrow$1783 &\bfseries $\uparrow$1845 &\bfseries $\uparrow$1447 &\bfseries $\uparrow$1816 &\bfseries $\uparrow$682 &\bfseries $\uparrow$1890 &\bfseries $\uparrow$1916 \\ 
	SK &\bfseries $\uparrow$1105 &\bfseries $\uparrow$919 &\bfseries $\uparrow$1109 &\bfseries $\uparrow$954 &\bfseries $\uparrow$1142 & \itshape $\downarrow$746 &\bfseries $\uparrow$904 &\bfseries $\uparrow$970 &\bfseries $\uparrow$1084 &\bfseries $\uparrow$465 &\bfseries $\uparrow$937 &\bfseries $\uparrow$34 &\bfseries $\uparrow$1012 &\bfseries $\uparrow$1162 \\ 
 \end{tabularx}%
 }
 \caption{The impact of additional departures to Brexit}
\label{Additional Exits to Brexit}%
\end{footnotesize}
\end{table}
\addtocounter{table}{-1}
\begin{table}[!htbp]
  \centering
\begin{footnotesize}
\centerline{
\rowcolors{1}{gray!20}{}
\begin{tabularx}{1.15\textwidth}{l CCCCC CCCCC CCC} \toprule \hiderowcolors
 &IT &IE &LT &LU &LV &MT &NL &PL &PT & RO & SE & SI & SK\\  \hline
\showrowcolors
	AT & \itshape $\downarrow$11 &\bfseries $\uparrow$475 &\bfseries $\uparrow$572 &\bfseries $\uparrow$545 &\bfseries $\uparrow$561 &\bfseries $\uparrow$557 &\bfseries $\uparrow$189 &\bfseries $\uparrow$149 &\bfseries $\uparrow$523 &\bfseries $\uparrow$674 &\bfseries $\uparrow$305 &\bfseries $\uparrow$559 &\bfseries $\uparrow$1223 \\ 
	BE & \itshape $\downarrow$129 &\bfseries $\uparrow$304 &\bfseries $\uparrow$385 &\bfseries $\uparrow$349 &\bfseries $\uparrow$373 &\bfseries $\uparrow$361 &\bfseries $\uparrow$16 & \itshape $\downarrow$150 &\bfseries $\uparrow$355 &\bfseries $\uparrow$498 &\bfseries $\uparrow$148 &\bfseries $\uparrow$371 &\bfseries $\uparrow$2253 \\ 
	BG &\bfseries $\uparrow$243 &\bfseries $\uparrow$769 &\bfseries $\uparrow$769 &\bfseries $\uparrow$705 &\bfseries $\uparrow$742 &\bfseries $\uparrow$708 &\bfseries $\uparrow$427 &\bfseries $\uparrow$493 &\bfseries $\uparrow$775 &\bfseries $\uparrow$975 &\bfseries $\uparrow$538 &\bfseries $\uparrow$742 & \itshape $\downarrow$158 \\ 
	CY &\bfseries $\uparrow$1666 &\bfseries $\uparrow$2293 &\bfseries $\uparrow$2363 &\bfseries $\uparrow$2300 &\bfseries $\uparrow$2345 &\bfseries $\uparrow$2293 &\bfseries $\uparrow$2121 &\bfseries $\uparrow$3289 &\bfseries $\uparrow$2503 &\bfseries $\uparrow$2726 &\bfseries $\uparrow$2221 &\bfseries $\uparrow$2351 &\bfseries $\uparrow$2438 \\ 
	CZ & \itshape $\downarrow$95 &\bfseries $\uparrow$350 &\bfseries $\uparrow$428 &\bfseries $\uparrow$392 &\bfseries $\uparrow$416 &\bfseries $\uparrow$401 &\bfseries $\uparrow$54 & \itshape $\downarrow$40 &\bfseries $\uparrow$434 &\bfseries $\uparrow$547 &\bfseries $\uparrow$218 &\bfseries $\uparrow$414 &\bfseries $\uparrow$438 \\ 
	DE & \itshape $\downarrow$259 & \itshape $\downarrow$419 & \itshape $\downarrow$357 & \itshape $\downarrow$375 & \itshape $\downarrow$361 & \itshape $\downarrow$363 & \itshape $\downarrow$593 &\bfseries $\uparrow$42 & \itshape $\downarrow$318 & \itshape $\downarrow$112 & \itshape $\downarrow$504 & \itshape $\downarrow$369 & \itshape $\downarrow$340 \\ 
	DK &\bfseries $\uparrow$407 &\bfseries $\uparrow$899 &\bfseries $\uparrow$964 &\bfseries $\uparrow$906 &\bfseries $\uparrow$945 &\bfseries $\uparrow$916 &\bfseries $\uparrow$644 &\bfseries $\uparrow$918 &\bfseries $\uparrow$1068 &\bfseries $\uparrow$1179 &\bfseries $\uparrow$824 &\bfseries $\uparrow$934 &\bfseries $\uparrow$1012 \\ 
	EE &\bfseries $\uparrow$1439 &\bfseries $\uparrow$2103 &\bfseries $\uparrow$2125 &\bfseries $\uparrow$2059 &\bfseries $\uparrow$2091 &\bfseries $\uparrow$2076 &\bfseries $\uparrow$1851 &\bfseries $\uparrow$2872 &\bfseries $\uparrow$2206 &\bfseries $\uparrow$2504 &\bfseries $\uparrow$1945 &\bfseries $\uparrow$2095 &\bfseries $\uparrow$2203 \\ 
	EL & \itshape $\downarrow$116 &\bfseries $\uparrow$310 &\bfseries $\uparrow$395 &\bfseries $\uparrow$357 &\bfseries $\uparrow$393 &\bfseries $\uparrow$371 &\bfseries $\uparrow$26 & \itshape $\downarrow$117 &\bfseries $\uparrow$380 &\bfseries $\uparrow$517 &\bfseries $\uparrow$175 &\bfseries $\uparrow$392 &\bfseries $\uparrow$396 \\ 
	ES & \itshape $\downarrow$218 & \itshape $\downarrow$382 & \itshape $\downarrow$310 & \itshape $\downarrow$331 & \itshape $\downarrow$317 & \itshape $\downarrow$321 & \itshape $\downarrow$639 &\bfseries $\uparrow$1320 & \itshape $\downarrow$346 & \itshape $\downarrow$158 & \itshape $\downarrow$538 & \itshape $\downarrow$325 & \itshape $\downarrow$308 \\ 
	FI &\bfseries $\uparrow$415 &\bfseries $\uparrow$926 &\bfseries $\uparrow$988 &\bfseries $\uparrow$930 &\bfseries $\uparrow$960 &\bfseries $\uparrow$936 &\bfseries $\uparrow$669 &\bfseries $\uparrow$959 &\bfseries $\uparrow$1098 &\bfseries $\uparrow$1211 &\bfseries $\uparrow$855 &\bfseries $\uparrow$954 &\bfseries $\uparrow$1025 \\ 
	FR & \itshape $\downarrow$709 & \itshape $\downarrow$424 & \itshape $\downarrow$360 & \itshape $\downarrow$374 & \itshape $\downarrow$365 & \itshape $\downarrow$361 & \itshape $\downarrow$650 &\bfseries $\uparrow$124 & \itshape $\downarrow$361 & \itshape $\downarrow$148 & \itshape $\downarrow$551 & \itshape $\downarrow$373 & \itshape $\downarrow$352 \\ 
	HR &\bfseries $\uparrow$583 &\bfseries $\uparrow$1097 &\bfseries $\uparrow$1182 &\bfseries $\uparrow$1164 &\bfseries $\uparrow$1175 &\bfseries $\uparrow$1176 &\bfseries $\uparrow$948 &\bfseries $\uparrow$1344 &\bfseries $\uparrow$1229 &\bfseries $\uparrow$1459 &\bfseries $\uparrow$972 &\bfseries $\uparrow$1175 &\bfseries $\uparrow$574 \\ 
	HU & \itshape $\downarrow$73 &\bfseries $\uparrow$383 &\bfseries $\uparrow$458 &\bfseries $\uparrow$443 &\bfseries $\uparrow$453 &\bfseries $\uparrow$454 &\bfseries $\uparrow$89 &\bfseries $\uparrow$8 &\bfseries $\uparrow$424 &\bfseries $\uparrow$565 &\bfseries $\uparrow$214 &\bfseries $\uparrow$451 &\bfseries $\uparrow$477 \\ 
	IT &\cellcolor{gray!80} & \itshape $\downarrow$417 & \itshape $\downarrow$352 & \itshape $\downarrow$371 & \itshape $\downarrow$356 & \itshape $\downarrow$360 & \itshape $\downarrow$604 &\bfseries $\uparrow$326 & \itshape $\downarrow$327 & \itshape $\downarrow$105 & \itshape $\downarrow$511 & \itshape $\downarrow$363 & \itshape $\downarrow$336 \\ 
	IE &\bfseries $\uparrow$531 &\cellcolor{gray!80} &\bfseries $\uparrow$1086 &\bfseries $\uparrow$1065 &\bfseries $\uparrow$1103 &\bfseries $\uparrow$1083 &\bfseries $\uparrow$851 &\bfseries $\uparrow$1212 &\bfseries $\uparrow$1169 &\bfseries $\uparrow$1377 &\bfseries $\uparrow$915 &\bfseries $\uparrow$1104 &\bfseries $\uparrow$1163 \\ 
	LT &\bfseries $\uparrow$886 &\bfseries $\uparrow$1483 &\cellcolor{gray!80} &\bfseries $\uparrow$1503 &\bfseries $\uparrow$1553 &\bfseries $\uparrow$1512 &\bfseries $\uparrow$1377 &\bfseries $\uparrow$1846 &\bfseries $\uparrow$1567 &\bfseries $\uparrow$1890 &\bfseries $\uparrow$1295 &\bfseries $\uparrow$1557 &\bfseries $\uparrow$1557 \\ 
	LU &\bfseries $\uparrow$1834 &\bfseries $\uparrow$2447 &\bfseries $\uparrow$2537 &\cellcolor{gray!80} &\bfseries $\uparrow$2529 &\bfseries $\uparrow$2437 &\bfseries $\uparrow$2301 &\bfseries $\uparrow$3598 &\bfseries $\uparrow$2717 &\bfseries $\uparrow$2944 &\bfseries $\uparrow$2368 &\bfseries $\uparrow$2540 &\bfseries $\uparrow$2581 \\ 
	LV &\bfseries $\uparrow$1227 &\bfseries $\uparrow$1837 &\bfseries $\uparrow$1937 &\bfseries $\uparrow$1770 &\cellcolor{gray!80} &\bfseries $\uparrow$1792 &\bfseries $\uparrow$1609 &\bfseries $\uparrow$2377 &\bfseries $\uparrow$1912 &\bfseries $\uparrow$2312 &\bfseries $\uparrow$1652 &\bfseries $\uparrow$1876 &\bfseries $\uparrow$1949 \\ 
	MT &\bfseries $\uparrow$1913 &\bfseries $\uparrow$2560 &\bfseries $\uparrow$2645 &\bfseries $\uparrow$2532 &\bfseries $\uparrow$2621 &\cellcolor{gray!80} &\bfseries $\uparrow$2380 &\bfseries $\uparrow$3782 &\bfseries $\uparrow$2771 &\bfseries $\uparrow$3037 &\bfseries $\uparrow$2492 &\bfseries $\uparrow$2626 &\bfseries $\uparrow$2678 \\ 
	NL & \itshape $\downarrow$327 &\bfseries $\uparrow$71 &\bfseries $\uparrow$144 &\bfseries $\uparrow$92 &\bfseries $\uparrow$125 &\bfseries $\uparrow$104 &\cellcolor{gray!80} & \itshape $\downarrow$569 &\bfseries $\uparrow$65 &\bfseries $\uparrow$158 & \itshape $\downarrow$133 &\bfseries $\uparrow$121 &\bfseries $\uparrow$143 \\ 
	PL &\bfseries $\uparrow$705 & \itshape $\downarrow$463 & \itshape $\downarrow$373 & \itshape $\downarrow$373 & \itshape $\downarrow$365 & \itshape $\downarrow$364 & \itshape $\downarrow$867 &\cellcolor{gray!80} & \itshape $\downarrow$468 & \itshape $\downarrow$507 & \itshape $\downarrow$648 & \itshape $\downarrow$374 & \itshape $\downarrow$392 \\ 
	PT & \itshape $\downarrow$85 &\bfseries $\uparrow$361 &\bfseries $\uparrow$437 &\bfseries $\uparrow$400 &\bfseries $\uparrow$424 &\bfseries $\uparrow$412 &\bfseries $\uparrow$59 & \itshape $\downarrow$41 &\cellcolor{gray!80} &\bfseries $\uparrow$559 &\bfseries $\uparrow$230 &\bfseries $\uparrow$424 &\bfseries $\uparrow$438 \\ 
	RO & \itshape $\downarrow$368 & \itshape $\downarrow$63 &\bfseries $\uparrow$16 & \itshape $\downarrow$2 &\bfseries $\uparrow$14 &\bfseries $\uparrow$8 & \itshape $\downarrow$401 & \itshape $\downarrow$778 & \itshape $\downarrow$37 &\cellcolor{gray!80} & \itshape $\downarrow$235 &\bfseries $\uparrow$8 &\bfseries $\uparrow$8 \\ 
	SE & \itshape $\downarrow$64 &\bfseries $\uparrow$392 &\bfseries $\uparrow$467 &\bfseries $\uparrow$451 &\bfseries $\uparrow$464 &\bfseries $\uparrow$463 &\bfseries $\uparrow$101 &\bfseries $\uparrow$35 &\bfseries $\uparrow$438 &\bfseries $\uparrow$574 &\cellcolor{gray!80} &\bfseries $\uparrow$463 &\bfseries $\uparrow$479 \\ 
	SI &\bfseries $\uparrow$1189 &\bfseries $\uparrow$1792 &\bfseries $\uparrow$1892 &\bfseries $\uparrow$1743 &\bfseries $\uparrow$1858 &\bfseries $\uparrow$1755 &\bfseries $\uparrow$1584 &\bfseries $\uparrow$2329 &\bfseries $\uparrow$1879 &\bfseries $\uparrow$2253 &\bfseries $\uparrow$1626 &\cellcolor{gray!80} &\bfseries $\uparrow$1928 \\ 
	SK &\bfseries $\uparrow$430 &\bfseries $\uparrow$929 &\bfseries $\uparrow$991 &\bfseries $\uparrow$944 &\bfseries $\uparrow$974 &\bfseries $\uparrow$960 &\bfseries $\uparrow$692 &\bfseries $\uparrow$988 &\bfseries $\uparrow$1103 &\bfseries $\uparrow$1223 &\bfseries $\uparrow$880 &\bfseries $\uparrow$972 &\cellcolor{gray!80} \\ 
   \end{tabularx}%
}
\caption{The impact of additional departures to Brexit (continued)}
  \label{Additional Exits to Brexit 2}%
\end{footnotesize}
\end{table}%

%% file: Appendix_D_alternative.tex
\begin{table}[!htbp]

\centering

\begin{footnotesize}
  
\centerline{
\rowcolors{1}{gray!20}{}
\begin{tabularx}{1.15\textwidth}{l CCCCC CCCCC CCCC} 
\toprule \hiderowcolors
&AT &BE &BG &CY &CZ &DE &DK &EE &EL & ES & FI & FR & HU & IT \\ \hline 
\showrowcolors
    AT    &\cellcolor{gray!80} &  \bfseries $\uparrow$584   &  \bfseries $\uparrow$758   &  \bfseries $\uparrow$648   &  \bfseries $\uparrow$756   &\itshape $\downarrow$553  &  \bfseries $\uparrow$573   &  \bfseries $\uparrow$654   &  \bfseries $\uparrow$730   &  \bfseries $\uparrow$424   &  \bfseries $\uparrow$605   &\itshape $\downarrow$181  &  \bfseries $\uparrow$773   &  \bfseries $\uparrow$204 \\ 
    BE    &  \bfseries $\uparrow$454   &\cellcolor{gray!80} &  \bfseries $\uparrow$512   &  \bfseries $\uparrow$411   &  \bfseries $\uparrow$514   &\itshape $\downarrow$719  &  \bfseries $\uparrow$332   &  \bfseries $\uparrow$414   &  \bfseries $\uparrow$481   &  \bfseries $\uparrow$51    &  \bfseries $\uparrow$365   &\itshape $\downarrow$455  &  \bfseries $\uparrow$535   &\itshape $\downarrow$108 \\ 
    BG    &  \bfseries $\uparrow$860   &  \bfseries $\uparrow$762   &\cellcolor{gray!80} &  \bfseries $\uparrow$797   &  \bfseries $\uparrow$954   &\itshape $\downarrow$413  &  \bfseries $\uparrow$746   &  \bfseries $\uparrow$800   &  \bfseries $\uparrow$914   &  \bfseries $\uparrow$597   &  \bfseries $\uparrow$778   &\itshape $\downarrow$45   &  \bfseries $\uparrow$959   &  \bfseries $\uparrow$304 \\ 
    CY    &  \bfseries $\uparrow$2648  &  \bfseries $\uparrow$2542  &  \bfseries $\uparrow$2733  &\cellcolor{gray!80} &  \bfseries $\uparrow$2770  &  \bfseries $\uparrow$659   &  \bfseries $\uparrow$2509  &  \bfseries $\uparrow$2531  &  \bfseries $\uparrow$2714  &  \bfseries $\uparrow$2750  &  \bfseries $\uparrow$2537  &  \bfseries $\uparrow$1530  &  \bfseries $\uparrow$2775  &  \bfseries $\uparrow$2081 \\ 
    CZ    &  \bfseries $\uparrow$515   &  \bfseries $\uparrow$410   &  \bfseries $\uparrow$591   &  \bfseries $\uparrow$486   &\cellcolor{gray!80} &\itshape $\downarrow$670  &  \bfseries $\uparrow$407   &  \bfseries $\uparrow$489   &  \bfseries $\uparrow$560   &  \bfseries $\uparrow$157   &  \bfseries $\uparrow$440   &\itshape $\downarrow$380  &  \bfseries $\uparrow$608   &  \bfseries $\uparrow$0 \\ 
    DE    &\itshape $\downarrow$373  &\itshape $\downarrow$498  &\itshape $\downarrow$301  &\itshape $\downarrow$349  &\itshape $\downarrow$332  &\cellcolor{gray!80} &\itshape $\downarrow$459  &\itshape $\downarrow$350  &\itshape $\downarrow$363  &\itshape $\downarrow$186  &\itshape $\downarrow$425  &\itshape $\downarrow$422  &\itshape $\downarrow$308  &\itshape $\downarrow$131 \\ 
    DK    &  \bfseries $\uparrow$1079  &  \bfseries $\uparrow$973   &  \bfseries $\uparrow$1149  &  \bfseries $\uparrow$1023  &  \bfseries $\uparrow$1159  &\itshape $\downarrow$295  &\cellcolor{gray!80} &  \bfseries $\uparrow$1026  &  \bfseries $\uparrow$1125  &  \bfseries $\uparrow$940   &  \bfseries $\uparrow$984   &  \bfseries $\uparrow$136   &  \bfseries $\uparrow$1179  &  \bfseries $\uparrow$552 \\ 
    EE    &  \bfseries $\uparrow$2411  &  \bfseries $\uparrow$2317  &  \bfseries $\uparrow$2492  &  \bfseries $\uparrow$2297  &  \bfseries $\uparrow$2515  &  \bfseries $\uparrow$525   &  \bfseries $\uparrow$2292  &\cellcolor{gray!80} &  \bfseries $\uparrow$2480  &  \bfseries $\uparrow$2442  &  \bfseries $\uparrow$2327  &  \bfseries $\uparrow$1337  &  \bfseries $\uparrow$2540  &  \bfseries $\uparrow$1854 \\ 
    EL    &  \bfseries $\uparrow$470   &  \bfseries $\uparrow$360   &  \bfseries $\uparrow$534   &  \bfseries $\uparrow$437   &  \bfseries $\uparrow$536   &\itshape $\downarrow$702  &  \bfseries $\uparrow$364   &  \bfseries $\uparrow$438   &\cellcolor{gray!80} &  \bfseries $\uparrow$79    &  \bfseries $\uparrow$398   &\itshape $\downarrow$432  &  \bfseries $\uparrow$564   &\itshape $\downarrow$76 \\ 
    ES    &\itshape $\downarrow$372  &\itshape $\downarrow$521  &\itshape $\downarrow$295  &\itshape $\downarrow$314  &\itshape $\downarrow$343  &\itshape $\downarrow$661  &\itshape $\downarrow$449  &\itshape $\downarrow$312  &\itshape $\downarrow$384  &\cellcolor{gray!80} &\itshape $\downarrow$416  &\itshape $\downarrow$538  &\itshape $\downarrow$313  &\itshape $\downarrow$193 \\ 
    FI    &  \bfseries $\uparrow$1092  &  \bfseries $\uparrow$980   &  \bfseries $\uparrow$1165  &  \bfseries $\uparrow$1038  &  \bfseries $\uparrow$1171  &\itshape $\downarrow$291  &  \bfseries $\uparrow$961   &  \bfseries $\uparrow$1045  &  \bfseries $\uparrow$1138  &  \bfseries $\uparrow$962   &\cellcolor{gray!80} &  \bfseries $\uparrow$154   &  \bfseries $\uparrow$1195  &  \bfseries $\uparrow$556 \\ 
    FR    &\itshape $\downarrow$384  &\itshape $\downarrow$528  &\itshape $\downarrow$318  &\itshape $\downarrow$358  &\itshape $\downarrow$355  &\itshape $\downarrow$554  &\itshape $\downarrow$477  &\itshape $\downarrow$359  &\itshape $\downarrow$391  &\itshape $\downarrow$227  &\itshape $\downarrow$443  &\cellcolor{gray!80} &\itshape $\downarrow$326  &\itshape $\downarrow$290 \\ 
    HU    &  \bfseries $\uparrow$572   &  \bfseries $\uparrow$460   &  \bfseries $\uparrow$628   &  \bfseries $\uparrow$514   &  \bfseries $\uparrow$637   &\itshape $\downarrow$646  &  \bfseries $\uparrow$455   &  \bfseries $\uparrow$520   &  \bfseries $\uparrow$606   &  \bfseries $\uparrow$228   &  \bfseries $\uparrow$487   &\itshape $\downarrow$318  &\cellcolor{gray!80} &  \bfseries $\uparrow$79 \\ 
    IT    &\itshape $\downarrow$368  &\itshape $\downarrow$506  &\itshape $\downarrow$299  &\itshape $\downarrow$346  &\itshape $\downarrow$335  &\itshape $\downarrow$634  &\itshape $\downarrow$461  &\itshape $\downarrow$346  &\itshape $\downarrow$371  &\itshape $\downarrow$177  &\itshape $\downarrow$428  &\itshape $\downarrow$619  &\itshape $\downarrow$310  &\cellcolor{gray!80} \\ 
    IE    &  \bfseries $\uparrow$1241  &  \bfseries $\uparrow$1118  &  \bfseries $\uparrow$1308  &  \bfseries $\uparrow$1195  &  \bfseries $\uparrow$1316  &\itshape $\downarrow$198  &  \bfseries $\uparrow$1090  &  \bfseries $\uparrow$1215  &  \bfseries $\uparrow$1279  &  \bfseries $\uparrow$1180  &  \bfseries $\uparrow$1120  &  \bfseries $\uparrow$285   &  \bfseries $\uparrow$1350  &  \bfseries $\uparrow$718 \\ 
    LT    &  \bfseries $\uparrow$1687  &  \bfseries $\uparrow$1574  &  \bfseries $\uparrow$1799  &  \bfseries $\uparrow$1665  &  \bfseries $\uparrow$1778  &  \bfseries $\uparrow$102   &  \bfseries $\uparrow$1599  &  \bfseries $\uparrow$1680  &  \bfseries $\uparrow$1743  &  \bfseries $\uparrow$1746  &  \bfseries $\uparrow$1632  &  \bfseries $\uparrow$722   &  \bfseries $\uparrow$1806  &  \bfseries $\uparrow$1187 \\ 
    LU    &  \bfseries $\uparrow$2836  &  \bfseries $\uparrow$2758  &  \bfseries $\uparrow$2925  &  \bfseries $\uparrow$2699  &  \bfseries $\uparrow$2968  &  \bfseries $\uparrow$771   &  \bfseries $\uparrow$2715  &  \bfseries $\uparrow$2716  &  \bfseries $\uparrow$2905  &  \bfseries $\uparrow$2976  &  \bfseries $\uparrow$2742  &  \bfseries $\uparrow$1656  &  \bfseries $\uparrow$2972  &  \bfseries $\uparrow$2285 \\ 
    LV    &  \bfseries $\uparrow$2076  &  \bfseries $\uparrow$1969  &  \bfseries $\uparrow$2201  &  \bfseries $\uparrow$1973  &  \bfseries $\uparrow$2183  &  \bfseries $\uparrow$327   &  \bfseries $\uparrow$1967  &  \bfseries $\uparrow$2000  &  \bfseries $\uparrow$2143  &  \bfseries $\uparrow$2091  &  \bfseries $\uparrow$1999  &  \bfseries $\uparrow$1055  &  \bfseries $\uparrow$2206  &  \bfseries $\uparrow$1503 \\ 
    MT    &  \bfseries $\uparrow$2975  &  \bfseries $\uparrow$2867  &  \bfseries $\uparrow$3066  &  \bfseries $\uparrow$2808  &  \bfseries $\uparrow$3090  &  \bfseries $\uparrow$831   &  \bfseries $\uparrow$2821  &  \bfseries $\uparrow$2841  &  \bfseries $\uparrow$3042  &  \bfseries $\uparrow$3139  &  \bfseries $\uparrow$2854  &  \bfseries $\uparrow$1776  &  \bfseries $\uparrow$3097  &  \bfseries $\uparrow$2407 \\ 
    NL    &  \bfseries $\uparrow$205   &  \bfseries $\uparrow$98    &  \bfseries $\uparrow$265   &  \bfseries $\uparrow$137   &  \bfseries $\uparrow$264   &\itshape $\downarrow$965  &  \bfseries $\uparrow$69    &  \bfseries $\uparrow$136   &  \bfseries $\uparrow$239   &\itshape $\downarrow$408  &  \bfseries $\uparrow$104   &\itshape $\downarrow$806  &  \bfseries $\uparrow$287   &\itshape $\downarrow$538 \\ 
    PL    &\itshape $\downarrow$226  &\itshape $\downarrow$364  &\itshape $\downarrow$131  &\itshape $\downarrow$180  &\itshape $\downarrow$186  &\itshape $\downarrow$1278 &\itshape $\downarrow$285  &\itshape $\downarrow$175  &\itshape $\downarrow$224  &\itshape $\downarrow$1592 &\itshape $\downarrow$254  &\itshape $\downarrow$1177 &\itshape $\downarrow$159  &\itshape $\downarrow$899 \\ 
    PT    &  \bfseries $\uparrow$528   &  \bfseries $\uparrow$425   &  \bfseries $\uparrow$603   &  \bfseries $\uparrow$480   &  \bfseries $\uparrow$594   &\itshape $\downarrow$661  &  \bfseries $\uparrow$411   &  \bfseries $\uparrow$502   &  \bfseries $\uparrow$569   &  \bfseries $\uparrow$176   &  \bfseries $\uparrow$445   &\itshape $\downarrow$367  &  \bfseries $\uparrow$620   &  \bfseries $\uparrow$15 \\ 
    RO    &  \bfseries $\uparrow$45    &\itshape $\downarrow$49   &  \bfseries $\uparrow$112   &  \bfseries $\uparrow$33    &  \bfseries $\uparrow$117   &\itshape $\downarrow$1180 &\itshape $\downarrow$56   &  \bfseries $\uparrow$34    &  \bfseries $\uparrow$88    &\itshape $\downarrow$670  &\itshape $\downarrow$23   &\itshape $\downarrow$1042 &  \bfseries $\uparrow$131   &\itshape $\downarrow$785 \\ 
    SE    &  \bfseries $\uparrow$586   &  \bfseries $\uparrow$474   &  \bfseries $\uparrow$636   &  \bfseries $\uparrow$531   &  \bfseries $\uparrow$652   &\itshape $\downarrow$644  &  \bfseries $\uparrow$472   &  \bfseries $\uparrow$535   &  \bfseries $\uparrow$618   &  \bfseries $\uparrow$255   &  \bfseries $\uparrow$504   &\itshape $\downarrow$296  &  \bfseries $\uparrow$681   &  \bfseries $\uparrow$46 \\ 
    SI    &  \bfseries $\uparrow$2021  &  \bfseries $\uparrow$1914  &  \bfseries $\uparrow$2150  &  \bfseries $\uparrow$1936  &  \bfseries $\uparrow$2136  &  \bfseries $\uparrow$298   &  \bfseries $\uparrow$1926  &  \bfseries $\uparrow$1963  &  \bfseries $\uparrow$2085  &  \bfseries $\uparrow$2044  &  \bfseries $\uparrow$1960  &  \bfseries $\uparrow$1016  &  \bfseries $\uparrow$2154  &  \bfseries $\uparrow$1454 \\ 
    SK    &  \bfseries $\uparrow$1113  &  \bfseries $\uparrow$1000  &  \bfseries $\uparrow$1177  &  \bfseries $\uparrow$1056  &  \bfseries $\uparrow$1187  &\itshape $\downarrow$283  &  \bfseries $\uparrow$975   &  \bfseries $\uparrow$1063  &  \bfseries $\uparrow$1149  &  \bfseries $\uparrow$991   &  \bfseries $\uparrow$1006  &  \bfseries $\uparrow$163   &  \bfseries $\uparrow$1219  &  \bfseries $\uparrow$575 \\ 
    UK    &\itshape $\downarrow$398  &\itshape $\downarrow$533  &\itshape $\downarrow$318  &\itshape $\downarrow$355  &\itshape $\downarrow$364  &\itshape $\downarrow$579  &\itshape $\downarrow$473  &\itshape $\downarrow$358  &\itshape $\downarrow$396  &\itshape $\downarrow$196  &\itshape $\downarrow$439  &\itshape $\downarrow$594  &\itshape $\downarrow$336  &\itshape $\downarrow$277 \\ 
\end{tabularx}%
}    
\end{footnotesize}
\caption{The impact of any member state leaving before the accession of Croatia}
\label{before croatia 1}%
\end{table}%
\addtocounter{table}{-1}
\begin{table}[!htbp]
  \centering
\begin{footnotesize}

\centerline{
\rowcolors{1}{gray!20}{}
\begin{tabularx}{1.15\textwidth}{l CCCC CCCCC CCCC} \toprule \hiderowcolors
	 &IE & LT &  LU & LV &MT & NL & PL & PT & RO & SE &SI & SK &UK \\ \hline 
\showrowcolors
\hline
    AT    &  \bfseries $\uparrow$595   &  \bfseries $\uparrow$684   &  \bfseries $\uparrow$639   &  \bfseries $\uparrow$671   &  \bfseries $\uparrow$642   &  \bfseries $\uparrow$513   &  \bfseries $\uparrow$1123  &  \bfseries $\uparrow$725   &  \bfseries $\uparrow$100   &  \bfseries $\uparrow$520   &  \bfseries $\uparrow$665   &  \bfseries $\uparrow$757   &  \bfseries $\uparrow$523 \\ 
    BE    &  \bfseries $\uparrow$376   &  \bfseries $\uparrow$431   &  \bfseries $\uparrow$403   &  \bfseries $\uparrow$421   &  \bfseries $\uparrow$411   &  \bfseries $\uparrow$250   &  \bfseries $\uparrow$837   &  \bfseries $\uparrow$483   &  \bfseries $\uparrow$80    &  \bfseries $\uparrow$286   &  \bfseries $\uparrow$417   &  \bfseries $\uparrow$515   &  \bfseries $\uparrow$225 \\ 
    BG    &  \bfseries $\uparrow$772   &  \bfseries $\uparrow$838   &  \bfseries $\uparrow$780   &  \bfseries $\uparrow$815   &  \bfseries $\uparrow$793   &  \bfseries $\uparrow$664   &  \bfseries $\uparrow$1343  &  \bfseries $\uparrow$918   &  \bfseries $\uparrow$120   &  \bfseries $\uparrow$700   &  \bfseries $\uparrow$810   &  \bfseries $\uparrow$929   &  \bfseries $\uparrow$764 \\ 
    CY    &  \bfseries $\uparrow$2527  &  \bfseries $\uparrow$2587  &  \bfseries $\uparrow$2501  &  \bfseries $\uparrow$2575  &  \bfseries $\uparrow$2514  &  \bfseries $\uparrow$2380  &  \bfseries $\uparrow$3176  &  \bfseries $\uparrow$2726  &  \bfseries $\uparrow$300   &  \bfseries $\uparrow$2482  &  \bfseries $\uparrow$2568  &  \bfseries $\uparrow$2735  &  \bfseries $\uparrow$2373 \\ 
    CZ    &  \bfseries $\uparrow$431   &  \bfseries $\uparrow$510   &  \bfseries $\uparrow$455   &  \bfseries $\uparrow$499   &  \bfseries $\uparrow$468   &  \bfseries $\uparrow$319   &  \bfseries $\uparrow$929   &  \bfseries $\uparrow$553   &  \bfseries $\uparrow$90    &  \bfseries $\uparrow$358   &  \bfseries $\uparrow$502   &  \bfseries $\uparrow$581   &  \bfseries $\uparrow$314 \\ 
    DE    &\itshape $\downarrow$413  &\itshape $\downarrow$343  &\itshape $\downarrow$359  &\itshape $\downarrow$352  &\itshape $\downarrow$350  &\itshape $\downarrow$618  &  \bfseries $\uparrow$111   &\itshape $\downarrow$357  &\itshape $\downarrow$16   &\itshape $\downarrow$536  &\itshape $\downarrow$353  &\itshape $\downarrow$293  &  \bfseries $\uparrow$234 \\ 
    DK    &  \bfseries $\uparrow$966   &  \bfseries $\uparrow$1044  &  \bfseries $\uparrow$1013  &  \bfseries $\uparrow$1047  &  \bfseries $\uparrow$1023  &  \bfseries $\uparrow$920   &  \bfseries $\uparrow$1584  &  \bfseries $\uparrow$1126  &  \bfseries $\uparrow$150   &  \bfseries $\uparrow$920   &  \bfseries $\uparrow$1030  &  \bfseries $\uparrow$1137  &  \bfseries $\uparrow$906 \\ 
    EE    &  \bfseries $\uparrow$2313  &  \bfseries $\uparrow$2367  &  \bfseries $\uparrow$2268  &  \bfseries $\uparrow$2346  &  \bfseries $\uparrow$2278  &  \bfseries $\uparrow$2164  &  \bfseries $\uparrow$2932  &  \bfseries $\uparrow$2471  &  \bfseries $\uparrow$280   &  \bfseries $\uparrow$2252  &  \bfseries $\uparrow$2327  &  \bfseries $\uparrow$2506  &  \bfseries $\uparrow$2182 \\ 
    EL    &  \bfseries $\uparrow$400   &  \bfseries $\uparrow$455   &  \bfseries $\uparrow$428   &  \bfseries $\uparrow$452   &  \bfseries $\uparrow$430   &  \bfseries $\uparrow$272   &  \bfseries $\uparrow$878   &  \bfseries $\uparrow$506   &  \bfseries $\uparrow$80    &  \bfseries $\uparrow$313   &  \bfseries $\uparrow$442   &  \bfseries $\uparrow$545   &  \bfseries $\uparrow$254 \\ 
    ES    &\itshape $\downarrow$399  &\itshape $\downarrow$314  &\itshape $\downarrow$324  &\itshape $\downarrow$315  &\itshape $\downarrow$312  &\itshape $\downarrow$726  &\itshape $\downarrow$920  &\itshape $\downarrow$368  &\itshape $\downarrow$39   &\itshape $\downarrow$538  &\itshape $\downarrow$319  &\itshape $\downarrow$281  &  \bfseries $\uparrow$152 \\ 
    FI    &  \bfseries $\uparrow$976   &  \bfseries $\uparrow$1052  &  \bfseries $\uparrow$1021  &  \bfseries $\uparrow$1059  &  \bfseries $\uparrow$1041  &  \bfseries $\uparrow$931   &  \bfseries $\uparrow$1582  &  \bfseries $\uparrow$1138  &  \bfseries $\uparrow$150   &  \bfseries $\uparrow$937   &  \bfseries $\uparrow$1046  &  \bfseries $\uparrow$1148  &  \bfseries $\uparrow$912 \\ 
    FR    &\itshape $\downarrow$427  &\itshape $\downarrow$354  &\itshape $\downarrow$367  &\itshape $\downarrow$362  &\itshape $\downarrow$357  &\itshape $\downarrow$691  &\itshape $\downarrow$32   &\itshape $\downarrow$380  &\itshape $\downarrow$27   &\itshape $\downarrow$554  &\itshape $\downarrow$365  &\itshape $\downarrow$311  &  \bfseries $\uparrow$71 \\ 
    HU    &  \bfseries $\uparrow$487   &  \bfseries $\uparrow$558   &  \bfseries $\uparrow$507   &  \bfseries $\uparrow$542   &  \bfseries $\uparrow$520   &  \bfseries $\uparrow$368   &  \bfseries $\uparrow$960   &  \bfseries $\uparrow$607   &  \bfseries $\uparrow$90    &  \bfseries $\uparrow$409   &  \bfseries $\uparrow$532   &  \bfseries $\uparrow$636   &  \bfseries $\uparrow$369 \\ 
    IT    &\itshape $\downarrow$418  &\itshape $\downarrow$344  &\itshape $\downarrow$355  &\itshape $\downarrow$349  &\itshape $\downarrow$344  &\itshape $\downarrow$670  &\itshape $\downarrow$105  &\itshape $\downarrow$359  &\itshape $\downarrow$26   &\itshape $\downarrow$536  &\itshape $\downarrow$351  &\itshape $\downarrow$298  &  \bfseries $\uparrow$50 \\ 
    IE    &\cellcolor{gray!80} &  \bfseries $\uparrow$1242  &  \bfseries $\uparrow$1182  &  \bfseries $\uparrow$1240  &  \bfseries $\uparrow$1202  &  \bfseries $\uparrow$1074  &  \bfseries $\uparrow$1700  &  \bfseries $\uparrow$1280  &  \bfseries $\uparrow$160   &  \bfseries $\uparrow$1082  &  \bfseries $\uparrow$1228  &  \bfseries $\uparrow$1282  &  \bfseries $\uparrow$1038 \\ 
    LT    &  \bfseries $\uparrow$1636  &\cellcolor{gray!80} &  \bfseries $\uparrow$1640  &  \bfseries $\uparrow$1711  &  \bfseries $\uparrow$1659  &  \bfseries $\uparrow$1510  &  \bfseries $\uparrow$2168  &  \bfseries $\uparrow$1738  &  \bfseries $\uparrow$210   &  \bfseries $\uparrow$1532  &  \bfseries $\uparrow$1703  &  \bfseries $\uparrow$1805  &  \bfseries $\uparrow$1530 \\ 
    LU    &  \bfseries $\uparrow$2701  &  \bfseries $\uparrow$2789  &\cellcolor{gray!80} &  \bfseries $\uparrow$2771  &  \bfseries $\uparrow$2689  &  \bfseries $\uparrow$2567  &  \bfseries $\uparrow$3354  &  \bfseries $\uparrow$2919  &  \bfseries $\uparrow$330   &  \bfseries $\uparrow$2665  &  \bfseries $\uparrow$2746  &  \bfseries $\uparrow$2926  &  \bfseries $\uparrow$2556 \\ 
    LV    &  \bfseries $\uparrow$1979  &  \bfseries $\uparrow$2076  &  \bfseries $\uparrow$1952  &\cellcolor{gray!80} &  \bfseries $\uparrow$1972  &  \bfseries $\uparrow$1830  &  \bfseries $\uparrow$2586  &  \bfseries $\uparrow$2137  &  \bfseries $\uparrow$250   &  \bfseries $\uparrow$1917  &  \bfseries $\uparrow$2024  &  \bfseries $\uparrow$2172  &  \bfseries $\uparrow$1886 \\ 
    MT    &  \bfseries $\uparrow$2821  &  \bfseries $\uparrow$2885  &  \bfseries $\uparrow$2787  &  \bfseries $\uparrow$2894  &\cellcolor{gray!80} &  \bfseries $\uparrow$2708  &  \bfseries $\uparrow$3480  &  \bfseries $\uparrow$3038  &  \bfseries $\uparrow$340   &  \bfseries $\uparrow$2795  &  \bfseries $\uparrow$2874  &  \bfseries $\uparrow$3042  &  \bfseries $\uparrow$2686 \\ 
    NL    &  \bfseries $\uparrow$105   &  \bfseries $\uparrow$153   &  \bfseries $\uparrow$125   &  \bfseries $\uparrow$137   &  \bfseries $\uparrow$135   & \cellcolor{gray!80} &  \bfseries $\uparrow$453   &  \bfseries $\uparrow$235   &  \bfseries $\uparrow$60    &  \bfseries $\uparrow$44    &  \bfseries $\uparrow$131   &  \bfseries $\uparrow$242   &\itshape $\downarrow$139 \\ 
    PL    &\itshape $\downarrow$243  &\itshape $\downarrow$169  &\itshape $\downarrow$195  &\itshape $\downarrow$169  &\itshape $\downarrow$183  &\itshape $\downarrow$549  & \cellcolor{gray!80} &\itshape $\downarrow$214  &\itshape $\downarrow$17   &\itshape $\downarrow$387  &\itshape $\downarrow$176  &\itshape $\downarrow$116  &\itshape $\downarrow$544 \\ 
    PT    &  \bfseries $\uparrow$442   &  \bfseries $\uparrow$519   &  \bfseries $\uparrow$466   &  \bfseries $\uparrow$510   &  \bfseries $\uparrow$480   &  \bfseries $\uparrow$330   &  \bfseries $\uparrow$928   &\cellcolor{gray!80} &  \bfseries $\uparrow$90    &  \bfseries $\uparrow$370   &  \bfseries $\uparrow$502   &  \bfseries $\uparrow$593   &  \bfseries $\uparrow$328 \\ 
    RO    &\itshape $\downarrow$14   &  \bfseries $\uparrow$52    &  \bfseries $\uparrow$23    &  \bfseries $\uparrow$38    &  \bfseries $\uparrow$28    &\itshape $\downarrow$89   &  \bfseries $\uparrow$149   &  \bfseries $\uparrow$90    &\cellcolor{gray!80} &\itshape $\downarrow$108  &  \bfseries $\uparrow$38    &  \bfseries $\uparrow$115   &\itshape $\downarrow$405 \\ 
    SE    &  \bfseries $\uparrow$502   &  \bfseries $\uparrow$557   &  \bfseries $\uparrow$525   &  \bfseries $\uparrow$555   &  \bfseries $\uparrow$533   &  \bfseries $\uparrow$390   &  \bfseries $\uparrow$981   &  \bfseries $\uparrow$623   &  \bfseries $\uparrow$90    & \cellcolor{gray!80} &  \bfseries $\uparrow$546   &  \bfseries $\uparrow$655   &  \bfseries $\uparrow$374 \\ 
    SI    &  \bfseries $\uparrow$1935  &  \bfseries $\uparrow$2030  &  \bfseries $\uparrow$1912  &  \bfseries $\uparrow$2000  &  \bfseries $\uparrow$1927  &  \bfseries $\uparrow$1795  &  \bfseries $\uparrow$2539  &  \bfseries $\uparrow$2092  &  \bfseries $\uparrow$240   &  \bfseries $\uparrow$1877  & \cellcolor{gray!80} &  \bfseries $\uparrow$2133  &  \bfseries $\uparrow$1844 \\ 
    SK    &  \bfseries $\uparrow$980   &  \bfseries $\uparrow$1070  &  \bfseries $\uparrow$1042  &  \bfseries $\uparrow$1077  &  \bfseries $\uparrow$1055  &  \bfseries $\uparrow$953   &  \bfseries $\uparrow$1595  &  \bfseries $\uparrow$1152  &  \bfseries $\uparrow$150   &  \bfseries $\uparrow$949   &  \bfseries $\uparrow$1060  & \cellcolor{gray!80} &  \bfseries $\uparrow$926 \\ 
    UK    &\itshape $\downarrow$428  &\itshape $\downarrow$356  &\itshape $\downarrow$363  &\itshape $\downarrow$359  &\itshape $\downarrow$355  &\itshape $\downarrow$683  &\itshape $\downarrow$54   &\itshape $\downarrow$389  &\itshape $\downarrow$27   &\itshape $\downarrow$562  &\itshape $\downarrow$362  &\itshape $\downarrow$309  & \cellcolor{gray!80} \\ 
 \end{tabularx}%
}    

\end{footnotesize}
\caption{The impact of any member state leaving before the accession of Croatia (continued)}
  \label{before croatia 2}%
\end{table}